\documentclass[%
reprint,
superscriptaddress,
msmath,amssymb,
amsfonts,amsmath,floatfix
aps,
prl,
]{revtex4-1}
\pdfoutput=1
\usepackage{graphicx}
\usepackage{amsmath}
\usepackage{amssymb}
\usepackage{dsfont}
\usepackage{tensor}
\usepackage{color}
\usepackage{soul}
\usepackage{ulem}
\usepackage{cancel}
\usepackage{upgreek}
\usepackage{braket}
\usepackage{blindtext}
\usepackage{dsfont}
\usepackage{amsmath}
\usepackage{amssymb}
\usepackage{ulem}
\usepackage{float}
\usepackage[letterpaper,left=1.5cm,right=1.5cm,top=1.6 cm,bottom=2.5cm]{geometry}
\setstcolor{red}

\begin{document}
\pdfoutput=1
\preprint{APS/123-QED}

\title{Physics-informed neural networks for quantum control}

\author{Ariel Norambuena}
\email{ariel.norambuena@umayor.cl}
\affiliation{Centro de Optica e Información Cuántica, Universidad Mayor, camino la Piramide 5750, Huechuraba, Santiago, Chile}
\author{Marios Mattheakis}
\affiliation{John A. Paulson School of Engineering and Applied Sciences,
Harvard University, Cambridge, Massachusetts 02138, USA}
\author{Francisco J. Gonz\'alez}
\affiliation{Centro de Investigaci\'on DAiTA Lab, Facultad de Estudios Interdisciplinarios, Universidad Mayor, Chile}
\author{Ra\'ul Coto}
\email{raul.coto@protonmail.com}
\affiliation{Centro de Investigaci\'on DAiTA Lab, Facultad de Estudios Interdisciplinarios, Universidad Mayor, Chile}
\affiliation{Department of Physics, Florida International University, Miami, Florida 33199, USA}
\date{\today}

\begin{abstract}
Quantum control is a ubiquitous research field that has enabled physicists to delve into the dynamics and features of quantum systems, delivering powerful applications for various atomic, optical, mechanical, and solid-state systems. In recent years, traditional control techniques based on optimization processes have been translated into efficient artificial intelligence algorithms. Here, we introduce a computational method for optimal quantum control problems via physics-informed neural networks (PINNs). We apply our methodology to open quantum systems by efficiently solving the state-to-state transfer problem with high probabilities, short-time evolution, and using low-energy consumption controls. Furthermore, we illustrate the flexibility of PINNs to solve the same problem under changes in physical parameters and initial conditions, showing advantages in comparison with standard control techniques.
\end{abstract}
\maketitle

Optimal Quantum Control (QC) is crucial to exploit all the advantages of quantum systems, ranging from entangled states preparation and quantum registers to quantum sensing. Nowadays, QC can be achieved by means of controllable dissipative dynamics~\cite{Lin2013,Jamonneau2016}, measurement-induced backaction~\cite{Branczyk2007,Blok2014,Montenegro2017}, Lyapunov control~\cite{Bacciotti2001,Wang2021}, optimal pulse sequences~\cite{Cui2017}, and differentiable programming~\cite{Schafer,Coopmans}. These QC techniques serve multiple purposes including state preservation, state-to-state transfer~\cite{Cong_QControl}, dynamical decoupling in open systems~\cite{Viola1999,Viola2005,Zhen2016} and trajectory tracking~\cite{Liu2011,Wu2022}. Furthermore, we have witnessed powerful applications across multiple platforms, including atomic systems~\cite{Du14,Anderlini07}, light-matter systems~\cite{Calajo2019,Tancara2021}, solid-state devices~\cite{Zhou2017,Tian2019}, trapped ions~\cite{Leibfried2003}, among others. Dynamical QC stems from a time-dependent Hamiltonian that steers the dynamics~\cite{Ticozzi2012}, and it is subjected to several constraints like laser power, inhomogeneous frequency broadening, and relaxation processes, to name a few. Therefore, finding the optimal sequence for QC is highly cumbersome and generally depends on the system.

\begin{figure}[ht]
\centering
\includegraphics[width=1 \linewidth]{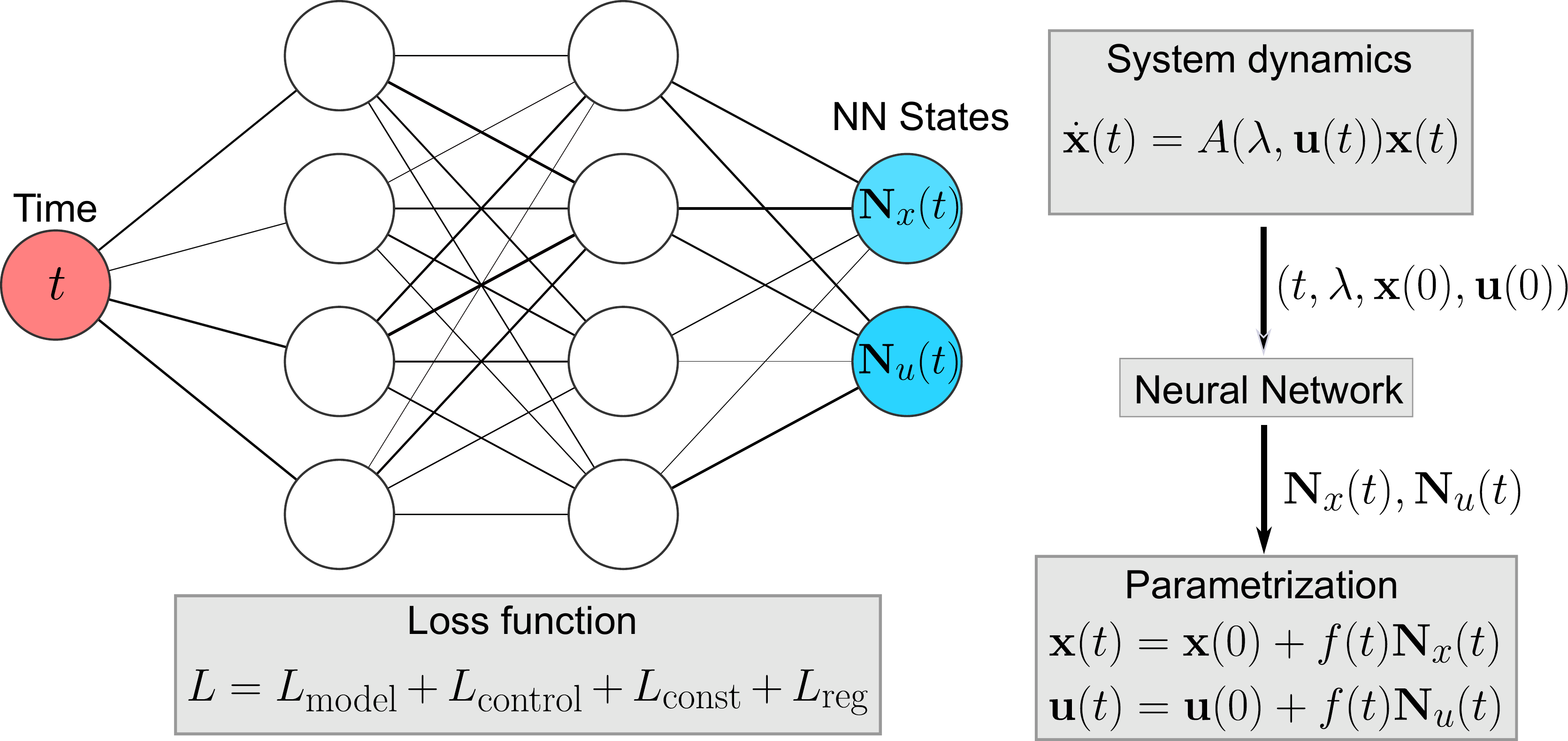}
\caption{PINNs architecture for solving quantum control problems. Quantum evolution can be translated into a dynamical system $\dot{\mathbf{x}}(t) = A(\mathbf{\lambda},\mathbf{u}(t)) \mathbf{x}(t)$, where $\mathbf{x}(t)$ and $\mathbf{u}(t)$ are the state and control vectors, respectively, and $\lambda$ are the system parameters. The input data (red circle) is given by the discrete time vector $t$, and the outputs of the neural network (NN) are $\mathbf{N}_x(t)$ and $\mathbf{N}_u(t)$ (blue circle). By minimizing the loss function $L$ the NN discover $\mathbf{N}_{x,u}(t)$ for the parameterized solutions $\mathbf{x}(t)$ and $\mathbf{u}(t)$.}
\label{fig:Figure1}
\end{figure}

Complex computational calculations are at the forefront of numerical methods to tackle down simulation of quantum systems. For instance, a parametrization of quantum states in terms of neural networks has enabled the approximation of many-body wavefunctions in closed quantum systems~\cite{Carleo2017,Cai2018,Choo2018}, and it has also been extended to approach the density operator in open dynamics (dissipative)~\cite{Torlai2018,Nagy2019,Yoshioka2019,Hartmann2019,Vicentini2019,Carrasquilla2021}. Along these ideas, other models have focused on hybrid implementations~\cite{Gardas2018,Liu2022,Koch2021}, probabilistic formulations based on positive operator-valued measure~\cite{Reh2021,Luo2022}, or data-driven model via time-averaged generators~\cite{Mazza2021}. Overall, estimating the dynamics of open quantum systems is a challenging problem. Here, machine learning provides versatile and promising algorithms to expand our alternatives towards completing this task~\cite{Niu2019,Brown21,Huang2022,Castaldo2021,Fosel2018,Zeng2020,Sivak2022}. However, combining time evolution and QC with artificial intelligence, being solved within a single deep learning method still needs to be explored. 

Neural Networks (NNs) are commonly trained with data allowing them to learn the dynamics of quantum systems. However, NNs that preserve the underlying physical laws without preliminary data would have practical advantages. Hence, physics-informed neural networks (PINNs) have been introduced as a new artificial intelligence paradigm that only requires the model itself~\cite{Raissi2019,Karniadakis2021}. This is a general physics-informed machine learning framework that has been applied to solve high-dimensional partial differential equations~\cite{Sirignano2018, Zeng2022}, many-body quantum systems \cite{Pfau2020,Zhu2023} and quantum fields~\cite{Martyn}, inverse problems using sparse and noisy data~\cite{Angeli2021}, and to discover underlying physics hidden in data structures~\cite{Shaan2021}. Since PINNs are coded using physical laws, they can be applied to any quantum evolution where the model is well known~\cite{Fang2019, Letellier2018,Hannachi1999,Tacchella2018,Denisov2002,Harnack2017,Kestner2022}. 

In this letter, we introduce a novel PINN architecture to find optimal control functions in open quantum systems. This is a data-free inverse modeling deep learning approach with a target dynamical behavior instead of data. Our approach suggests smooth control functions for driving quantum states to a pre-selected target state.

Let us consider the following $n$-dimensional non-autonomous dynamical system:
\begin{equation}
\dot{\mathbf{x}} = A(\mathbf{\lambda},\mathbf{u}(t)) \mathbf{x}(t), \quad \mathbf{x}(0) = \mathbf{x}_0, \; \quad \mathbf{u}(0) = \mathbf{u}_0,
\label{eq1}
\end{equation}
where $\mathbf{x}(t) = (x_1,...,x_n)^T \in \mathds{R}^n$, $\mathbf{u}(t) = (u_1,...,u_m)^T \in \mathds{R}^m$, and $\mathbf{\lambda} = (\lambda_1,...,\lambda_s)^T \in \mathds{R}^{s}$ are the state, control, and parameter vectors (with $n,m,s \geq 1$ and $m \leq n^2$), respectively. Here, $A(\mathbf{\lambda},\mathbf{u}(t))$ is a real $n \times n$ dynamical matrix that depends on $\mathbf{u}(t)$ (control) and $\mathbf{\lambda}$ (parameters). 

Given $\lambda$ and $\mathbf{x}(t)$ satisfying Eq.~\eqref{eq1}, we can apply machine learning to discover an optimal control vector $\mathbf{u}(t)$ such that the system evolves from $\mathbf{x}(0)$ to some desired target state $\mathbf{x}_d$. Techniques based on optimal control~\cite{Wendell1975}, Lyapunov control theory~\cite{Bacciotti2001} or linear control theory are based on optimization rules to find a suitable control vector $\mathbf{u}(t)$. Therefore, the main idea is to construct a PINN that minimizes a loss function to achieve optimal quantum control. 

\begin{figure}[ht!]
\centering
\includegraphics[width=1 \linewidth]{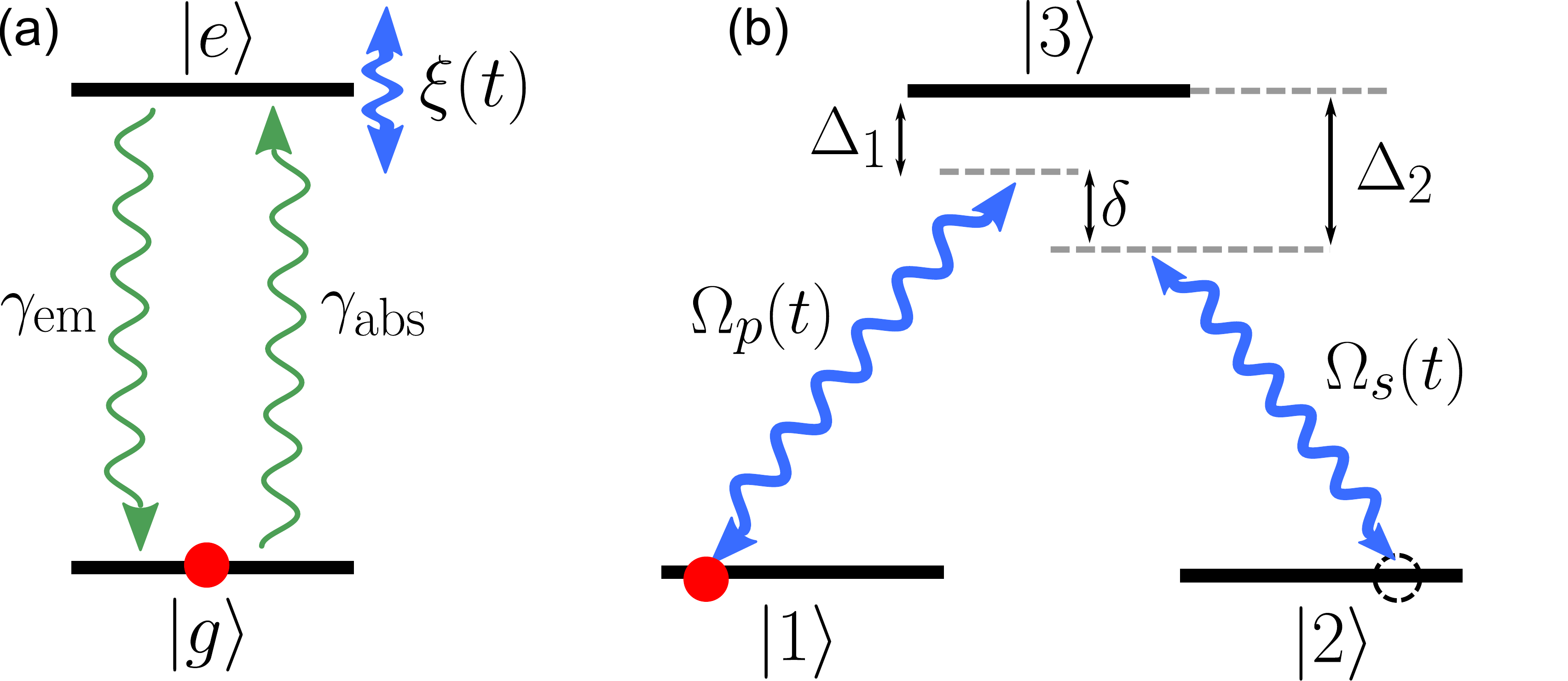}
\caption{(a) Open two-level system controlled by a time-dependent modulation $\xi(t)$. Losses are included through absorption ($\gamma_{\rm abs}$) and emission ($\gamma_{\rm em}$) processes. (b) Three-level $\Lambda$ system controlled by two driving fields $\Omega_p(t)$ and $\Omega_s(t)$. We include pure dephasing rates $\gamma_{i}$ ($i=1,2,3$) acting on each state $|i\rangle$. One- ($\Delta_{1,2}$) and two-photon ($\delta$) detunings are considered in our simulations.} 
\label{fig:Figure2}
\end{figure}

A feed forward NN is a powerful universal approximator for any vector function $F : \mathds{R}^p \mapsto \mathds{R}^{q}$ ($r, q \geq 1$)  (Universal Approximation Theorem)~\cite{Hornik1991}. Let us consider the NN architecture illustrated in Fig.~\ref{fig:Figure1} as a new paradigm to solve quantum control problems. We use an equally distributed time array $t = (t_1,...,t_{M})$ as the input to the NN, with $M$ representing the dimension of the sample points. PINNs do not require a structured mesh; thus, $t_i$ can be arbitrarily discretized. The NN consists of multiple hidden layers with activation function $\sin(\cdot)$ for the hidden neurons. This choice of activation has been shown to improve PINNs' performance in solving nonlinear dynamical systems~\cite{Mattheakis2020} and high-dimensional partial differential equations~\cite{Zeng2022}. The outputs of the NN are the solutions $\mathbf{N}_x(t) \in \mathds{R}^n$ and $\mathbf{N}_u(t) \in \mathds{R}^m$. We construct a neural state and control vectors that identically satisfy the initial conditions by using a hard constraint $\mathbf{x}(t) = \mathbf{x}(0)+f(t)\mathbf{N}_x(t)$ and $\mathbf{u}(t) = \mathbf{u}(0)+f(t)\mathbf{N}_u(t)$, where $f(t)= 1-e^{-t}$ is a function satisfying $f(0)=0$~\cite{Mattheakis2020}. This hard constraint approach avoids numerical errors in the initial conditions. The network parameters, weights and biases, are randomly initialized, and then, they are optimized  by minimizing a physics-informed loss function defined by 

\begin{equation} \label{LossFunction}
L = L_{\rm model} + L_{\rm control} + L_{\rm const} + L_{\rm reg}.
\end{equation}

The component $L_{\rm model}$ describes the system dynamics:

\begin{equation} \label{Loss_model}
L_{\rm model} = \sum_{i=1}^{M}\lvert\lvert \dot{\mathbf{x}}(t_i)-A(\mathbf{\lambda},\mathbf{u}(t_i))\mathbf{x}(t_i) \rvert\rvert^2,
\end{equation}

with $ \lvert\lvert \cdot \rvert\rvert$ representing a Euclidean distance. The time derivatives in the neural solutions are computed using the automatic differentiation method provided by PyTorch package~\cite{pytorch}. By minimizing the above functional, the state vector will approximately satisfy the system dynamics and, thus, the underlying physics. The second term on the right-hand side of Eq.~\eqref{LossFunction} represents the control, that reads
\begin{equation}
L_{\rm control} = \eta \sum_{i=1}^{M}\lvert\lvert \mathbf{x}(t_i)-\mathbf{x}_d \rvert\rvert^2, \quad 0 \leq \eta \leq 1,
\end{equation}

where the factor $\eta$ regulates the relevance of the control condition compared to the leading model component $L_\text{model}$. Note that $\mathbf{x}_d$ could be a constant (regulation) or time-dependent (trajectory tracking) vector depending on the control scheme. The term $L_{\rm const}$ could take into account additional physical constraints for the state/control vector, respectively, such as probability conservation or holonomic constraints of the form $H(\mathbf{x},\mathbf{u},t) = 0$. Finally, $L_{\rm reg}$ is a standard regularization loss term that encourages the network parameters to take relatively small values avoiding overfitting. We remark that imposing initial conditions into the loss function (soft constraint) is also possible, as illustrated in Ref.~\cite{Raissi2019}. The comparison between soft and hard constraints are discussed in Ref.~\cite{SM}. The effect of overfitting will be the prediction of a too complex ${\bf u}(t)$, which might be practically unfeasible for designing a real control. We introduce $L_{\rm reg}$ as a $l_2$-norm of the network weights $L_\text{reg}=\chi\sum_{i}\mbox{w}_i^2$, where $\chi$ is the regularization parameter.

The minimization of the loss function given in Eq.~\eqref{LossFunction} yields NN predictions that obey the underlying physics and suggest optimal control functions. For the training (minimization of Eq.~\eqref{LossFunction}), we employ Adam optimizer~\cite{adam2014}. Moreover, the points $t_i$ are randomly perturbed during training iteration--- this method has been shown to improve the training and the neural predictions~\cite{Sirignano2018, Mattheakis2020}. To highlight the method and keep the presentation elegant, we focused on low-dimensional Hilbert space examples. In the Supplemental Material, we demonstrate that the proposed PINN can successfully deal with larger systems.

\begin{figure}[t]
\centering
\includegraphics[width=0.8 \linewidth]{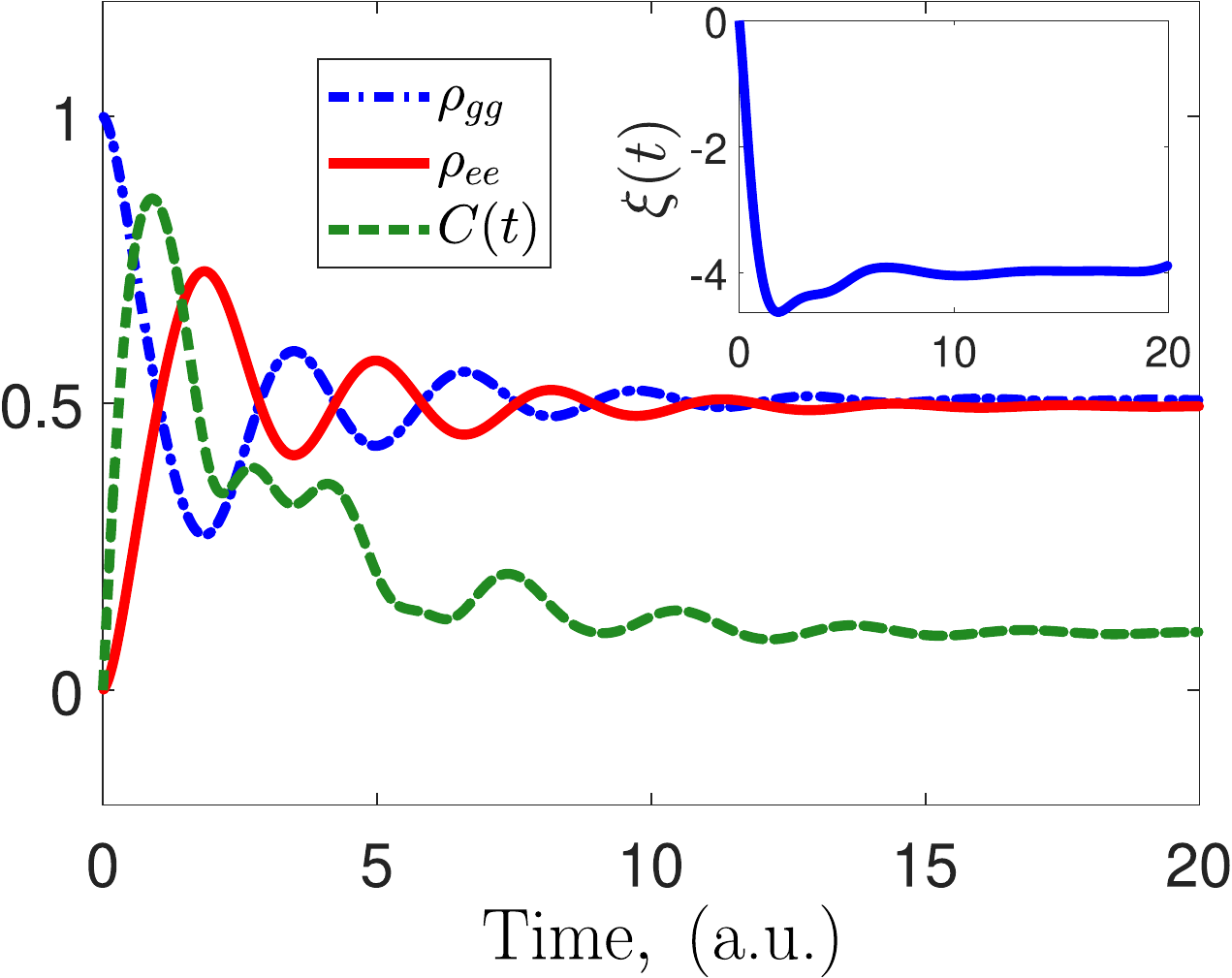}
\caption{Evolution of populations $\rho_{gg}(t), ~\rho_{ee}(t)$ and coherence $C(t)$ using the control function $\xi(t)$ predicted by the PINN. The architecture of the NN consists of $4$ hidden layers of $200$ neurons, it is trained for $4 \times 10^{4}$ epochs with a learning rate $10^{-4}$, $\chi = 10^{-3}$, and $\eta=1$.}
\label{fig:TLS}
\end{figure}

We consider a two-level system (TLS) as a proof-of-principle example to illustrate the use of PINNs for QC. We address the problem of generating Gibbs (mixed) states of the form $\rho_{\rm Gibbs} = Z^{-1}\sum_j e^{-\beta E_j}|j\rangle \langle j|$ with $Z = \mbox{Tr}(e^{-\beta H})$ (partition function) and $\beta = (k_B T)^{-1}$ (inverse temperature). The preparation of mixed states is relevant for simulating high-temperature superconductivity in variational quantum circuits~\cite{Sagastizabal2021}. In addition, QC of two-level systems is also relevant in the context of pulse reverse engineering~\cite{Ran2020}, feedback control~\cite{Branczyk2007}, optimal control theory~\cite{Wendell1975}, and controllable quantum dissipative dynamics~\cite{Wu2021}. Quantum transitions can be written in terms of the operators $\sigma_{ij} = |i\rangle\langle j|$ ($i,j = e,g$) being $|e\rangle$ ($|g\rangle$) the excited (ground) state. Let us consider the following Hamiltonian with a phase damping control
\begin{equation}
H(t) = \omega_z \sigma_z + \omega_x \sigma_x +  \xi(t) \sigma_{ee},
\end{equation}
with $\omega_{x,z}$ representing system parameters and $\xi(t)$ describing the unknown control field. Here, $\sigma_z = \sigma_{ee}-\sigma_{gg}$ and $\sigma_x = \sigma_{eg}+\sigma_{ge}$. Here, $\xi(t)$ is the control used to generate Gibbs states. To train our PINN we use the Markovian master equation $\dot{\rho} = -i[H(t),\rho] + \sum_{i=1,2} \gamma_i(L_i \rho L_i^{\dagger}-(1/2)\{ L_i^{\dagger}L_i,\rho \})$, with $[\cdot,\cdot]$ ($\{\cdot,\cdot\}$) representing the commutator (anticommutator). The amplitude damping channel is described by absorption ($L_1 = \sigma_{eg}$) and emission ($L_2 = \sigma_{ge}$) processes with rates $\gamma_1 = \gamma_{\rm abs}$ and $\gamma_2 = \gamma_{\rm em}$, respectively, see Fig.~\ref{fig:Figure2}-(a). In what follows, we use $\omega_z = 2$, $\omega_x = 1$, $\gamma_{\rm abs} = 0.1$, $\gamma_{\rm em} = 0.3$, and $\xi(0) = 0$. For $\xi(t)=0$, we get the steady-state (ss) $\rho^{\rm ss} = 0.2775|e\rangle\langle e| + 0.7225|g\rangle\langle g|+ [(-0.1106+i0.0083)|g\rangle \langle e|+c.c]$. Thus, we use $\xi(t)$ to drive the system to another ss, say, $\rho_d = (1/2)(|e\rangle \langle e|+|g\rangle \langle g|)$. Hence, we introduce the real state vector $\mathbf{x}(t) = (x_1,x_2,x_3,x_4)^T = (\rho_{gg}, \rho_{ee},\mbox{Re}(\rho_{eg}),\mbox{Im}(\rho_{eg}))^T$, where $\rho_{ij} = \langle i |\rho(t)|j\rangle$ are the elements of the density matrix. The dynamics can be written as $\dot{\mathbf{x}} = A(\lambda,\mathbf{u}(t))\mathbf{x}(t)$, with
\begin{equation}\label{DynamicalMatrix1}
A(\lambda,\mathbf{u}(t)) = \left(\begin{array}{cccc}
-\gamma_{\rm abs} & \gamma_{\rm em} & 0 & -2\omega_x \\
\gamma_{\rm abs} & -\gamma_{\rm em} & 0 & 2\omega_x  \\
0 & 0 & -\Gamma & -2\omega_z-\xi(t) \\  
\omega_x & -\omega_x & 2\omega_z+\xi(t) & -\Gamma
\end{array} \right),
\end{equation}
where $\lambda = (w_x,w_z,\gamma_{\rm abs},\gamma_{\rm em})$ is the set of parameters, $\mathbf{u}(t) = \xi(t)$ is the control vector that needs to be discovered, and $\Gamma = (1/2)(\gamma_1 + \gamma_2)$ is the effective dephasing rate. We remark that all results concerning the time evolution are simulated from a traditional ODE solver using the control obtained from the PINN. In Fig.~\ref{fig:TLS}, we plot the dynamics using the PINN's prediction for the control $\xi(t)$ (inset). The PINN discovers an optimal Gibbs state preservation with fidelity $F(\rho(t),\rho_d)=[\mbox{Tr}([\rho^{1/2}(t)\rho_d \rho^{1/2}(t)]^{1/2})]^2=0.99$ (for $t \geq 20$), and the steady-state approaches to $\rho_d$ within an error of $1\%$ for each component of the density matrix~\cite{SM}. We remark that our result outperforms the analytically optimized solution that finds $\rho_{gg}^{\rm ss} = 0.549$, $\rho_{ee}^{\rm ss} = 0.4510$, $\mbox{Re}(\rho_{eg}) = 0$, and $\mbox{Im}(\rho_{eg}) = 0.049$, for a constant control $\xi^{\rm ss} = -4$ (see~\cite{SM} for further details). The latter explains the asymptotic behavior for $\xi(t)$ predicted by the PINN. Moreover, we note in Fig.~\ref{fig:TLS} that the quantum coherence $C(t)=2|\rho_{eg}(t)|$ is highly activated during the transient dynamics in order to generate an equally distributed mixed state, but it asymptotically reaches $C\approx 0.1014$. 

We now focus on a $\Lambda$-configuration with two control fields (see  Fig.~\ref{fig:Figure2}-(b)), a platform for studying electromagnetically induced transparency~\cite{Boller1991,Fleischhauer2005}, coherent population trapping~\cite{Arimondo1976,Jyotsna1995}, and adiabatic population transfer~\cite{Kuklinski1989}. The latter has been dubbed Stimulated Raman Adiabatic Passage (STIRAP)~\cite{Bergmann1998}. Let's begin with the system Hamiltonian $H = \sum_i E_i \sigma_{ii} + H_c(t)$, where $E_i$ stands for the eigenenergies, $\sigma_{ij}=\ket{i}\bra{j}$, and $H_c(t)$ is the control Hamiltonian. In a multi-rotating frame and after the rotating wave approximation, the dynamics of the three-level system is governed by ($\hbar=1$):
\begin{equation}\label{H_S}
 H(t) = \delta \sigma_{22}+ \Delta_1 \sigma_{33}
 +\left(\frac{\Omega_{p}(t)}{2}\sigma_{31}+ \frac{\Omega_{s}(t)}{2}\sigma_{32}+ h.c. \right),
\end{equation}
where $\Delta_1 = E_3 - E_1-\omega_p$ and $\Delta_2 = E_3 - E_2-\omega_s$ are the one-photon detunings that originate from off-resonant driving fields with frequencies $\omega_p$ and $\omega_s$, while $\delta  = \Delta_1-\Delta_2$ is the two-photon detuning. Here, $\Omega_p(t)$ and $\Omega_s(t)$ are the control fields to be found. A similar Hamiltonian can be obtained from the interaction of a Nitrogen-Vacancy center with a Carbon-13 nuclear spin~\cite{Gonzalez,Coto17}. We aim to find the optimal control pulses ($\Omega_p(t)$ and $\Omega_s(t)$) to transfer population from state $|1\rangle$ to state $|2\rangle$ via the lossy intermediary state $|3\rangle$. Our goal is to train a PINN that completes the task reaching high fidelity, with (\textit{i}) minimizing the population in the state $|3\rangle$, and (\textit{ii}) minimizing the pulse area. We train our model using the Markovian master equation ($\hbar=1$)
\begin{equation}
\dot{\rho} = -i[H(t),\rho] + \sum_{i=1}^{3}\gamma_i(2\sigma_{ii}\rho\sigma_{ii}-\sigma_{ii}\rho -\rho\sigma_{ii}),
\end{equation}
where $\gamma_{i}>0$ are dephasing rates. We set $\gamma_3=0.14$ and $\gamma_1=\gamma_2=10^{-3}$ to account for larger dissipation in the excited state. We use the real vector $\mathbf{z}$ = $(\rho_{11}$,$\rho_{22}$,$\rho_{33}$,$\mbox{Re}[\rho_{12}]$,$\mbox{Im}[\rho_{12}]$,$\mbox{Re}[\rho_{13}]$,$\mbox{Im}[\rho_{13}]$,$\mbox{Re}[\rho_{23}]$,$\mbox{Im}[\rho_{23}])^T$ to write the dynamics (details are given in ~\cite{SM}). 

\begin{figure}[t]
\centering
\includegraphics[width=1 \linewidth]{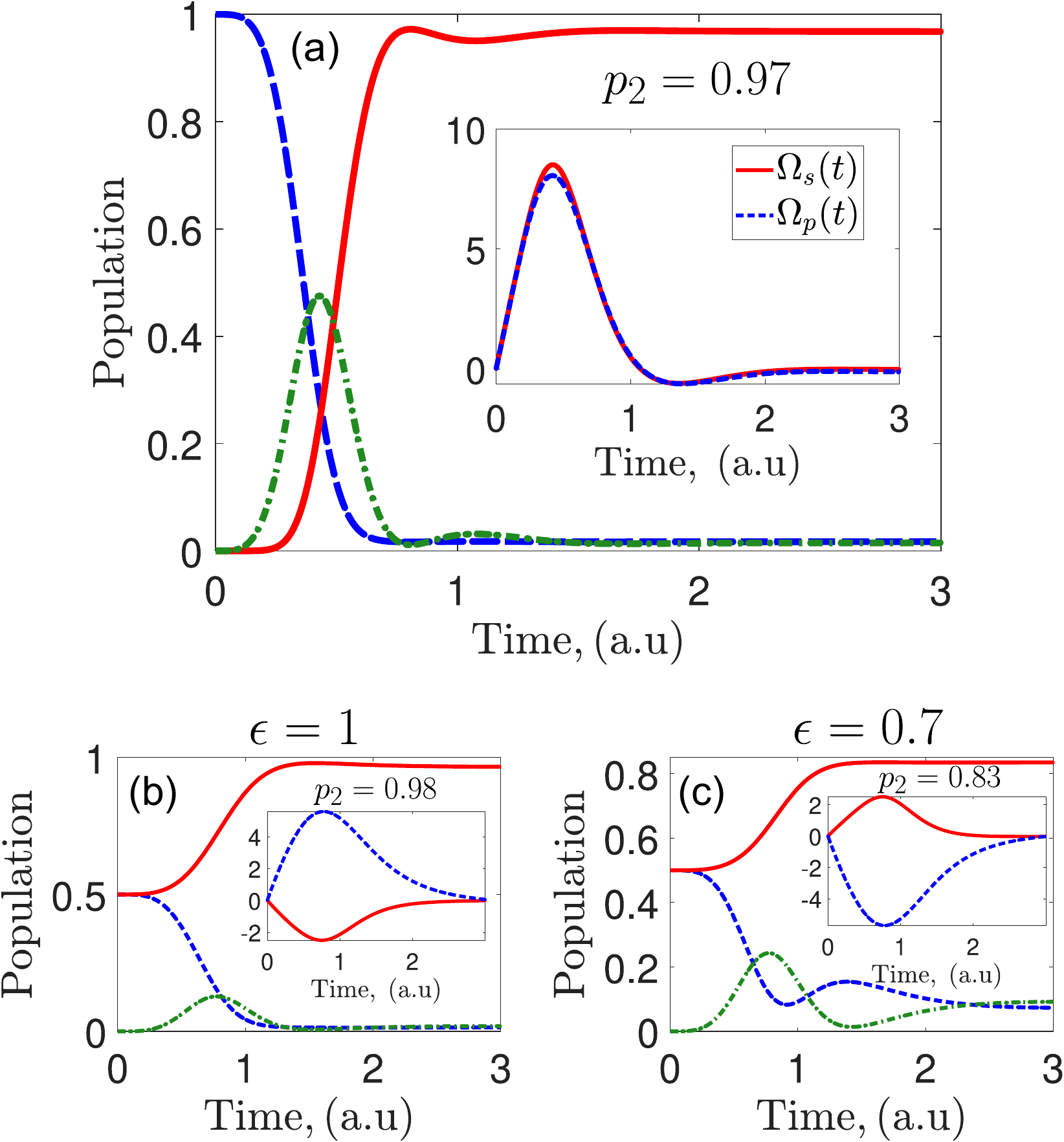}
\caption{(a) Population dynamics for the $\Lambda$-system using $\Omega_{s,p}(t)$ predicted by the PINN. (b) The PINN allows us to polarize the system starting from a coherent ($\epsilon=1$) and quasi-thermal state ($\epsilon=0.7$). The architecture of the NN includes $5$ hidden layers of $150$ neurons and it is optimized over $2\times10^{4}$ training epochs with learning rate $8\times10^{-3}$, $\eta=0.2$ and $\chi=2.8\times10^{-3}$~\cite{SM}.} 
\label{fig:NN_3level}
\end{figure}

In Fig.~\ref{fig:NN_3level}-(a), we show the population evolution and the predicted NN solutions for the control fields $\Omega_{s,p}(t)$. Note that our PINN successfully delivers a population transfer with smooth pulses. Furthermore, it attempts to implement a counterintuitive sequence, turning on the Stoke pulse $\Omega_s$ (red-solid) at the same time that the pump field $\Omega_p$ (blue-dashed)--- for a genuine counterintuitive sequence like STIRAP, Stokes pulse precedes the pump pulse. This is remarkable, considering that the PINN does not know QC theory or the relevance of following a dark state evolution. It is worthwhile noticing that the Stoke pulse shown in the inset of Fig.~\ref{fig:NN_3level}-(a) triggers the $|2\rangle \leftrightarrow |3\rangle$ transition, which in STIRAP serves the purpose of preparing a dark state~\cite{Coto17}, since initially, all the population is in state $|1\rangle$. 

For completeness, we consider a more challenging initial state given by $\rho(0)=\sigma_{11}/2 + \sigma_{22}/2 + \epsilon(\sigma_{12}+\sigma_{21})/2$. For $\epsilon=0$, we end up with a fully mixed state (without quantum coherence), while $\epsilon=1$ provides a balanced coherent state. Our PINN can handle this new task without changing the network's architecture, showing that PINNs provide a general and adaptive framework for inverse design (standard methods are not designed for this task). In Figs.~\ref{fig:NN_3level}-(b) and -(c), we show the population transfer and the corresponding pulse sequences for $\epsilon=1$ and $\epsilon=0.7$, respectively. Note that the PINN updates the pulses to deliver good polarizations. 

For a thorough benchmarking, we consider other control methods besides STIRAP~\cite{Kuklinski1989,Bergmann1998}, such as Stimulated Raman Exact Passage (STIREP)~\cite{Laforgue2022}, Inverse Engineering~\cite{chen2012,Lai1996} and Modified Superadiabatic Transitionless Driving (MOD-SATD)~\cite{Baksic2016,Zhou2017}. For detailed calculations of these pulses, see~\cite{SM}. In Table~\ref{Table1} we show a comparison. One can observe that PINN itself speeds up the population transfer with a high fidelity and using a small amount of energy. The transfer time $t_f$ is defined as the time required to reach the highest probability in the target state (more details in Ref.~\cite{SM}). The predicted control functions have the smallest area $\mathcal{A}$ compared to the other methods. We remark that the regularization $L_{\rm reg}$ penalizes the fields for being too large and provides smooth functions. Thus, we can control the amplitudes of the fields and the pulse area to achieve a less power-consuming transfer. Another important advantage of our protocol is the robustness against changes over initial training parameters. It is known that STIRAP deteriorates when increasing $\delta$~\cite{Romanenko1997,Vitanov2017}. Furthermore, the optimization for the other sequences with $\delta\neq 0$ is non trivial, and there is not much literature about it--- to our best knowledge. To compare the robustness of the earlier discovered pulse sequences, we also report in Table~\ref{Table1} the population transfer in the presence of two-photon detuning $\delta/2\pi=0.2$. We stress that no training or further optimizations have been made to account for the new $\delta$. Therefore, based on Table~\ref{Table1}, we conclude that: i) PINNs can reach high fidelities in a short time under low energy consumption, and ii) PINNs are very robust when initial training parameters are changed, delivering better results in comparison with standard methods. In~\cite{SM}, we show that PINNs can be easily trained to counteract the adversary effect of $\delta$.  

\begin{table}
\caption{\label{Table1}The Table shows the population $p_2=\mbox{Tr}[\rho\sigma_{22}]$, pulse area $\mathcal{A}=\int_0^{t_f} \, dt\sqrt{|\Omega_p(t)|^2+|\Omega_s(t)|^2}$ and transfer time $t_f$ (in arbitrary units). In parenthesis, we report the values with $\Delta_1/2\pi=0.2$ and $\delta/2\pi=0.2$, the one- and two-photon detuning, respectively. }
\begin{ruledtabular}
\begin{tabular}{cccccc}
  & \mbox{PINN}  & \mbox{STIRAP} & \mbox{STIREP} & \mbox{Inv. Eng.} & \mbox{MOD-SATD} \\
\hline
$p_{2}$ & $0.97 (0.93)$   & $0.98 (0.88)$ & $0.98 (0.91)$ & $0.97 (0.79)$ & $0.98 (0.89)$\\
$\mathcal{A}$ & $7.3$   & $128.6$ & $53.3$ & $19.8$ & $50.0$ \\
$t_f$ & $2.0$   & $35$ & $9.4$ & $3.0$ & $13$
\end{tabular}
\end{ruledtabular}
\end{table}

Finally, we extend our calculations to a four-level system and show that our PINN performs well against cross-talk to the newly added state, and also we check that our protocol deal with larger systems~\cite{SM}.

In this letter, we introduced a physics-informed neural network to find control functions in open quantum systems. We demonstrated a data-free deep learning approach that jointly solves the open dynamics of quantum systems and the inverse design of control functions. First, we applied this formalism to prepare a Gibbs state in a two-level system. Second, we applied it to state-to-state transfer in a three-level system. We found that the PINN provides a flexible method that adapts to different parameters, initial states, noise channels, and power consumption requirements. We hope that PINNs will be very attractive for problems such as adiabatic quantum computing, quantum gates, state preservation, manipulation in high-dimensional Hilbert spaces, initialization of entangled states, and time-dependent induced behavior in many-body systems.

A.N. and R.C acknowledge the financial support from the projects Fondecyt Iniciación $\#$11220266 and $\#$11180143, respectively.  F.J.G. acknowledges support from Universidad Mayor through the Doctoral fellowship.

\bibliographystyle{unsrt}

\newpage

\section{Supplemental Material for ``Physics-informed neural networks for quantum control''}

\begin{figure*}[t]
	\centering
	\includegraphics[scale=0.43]{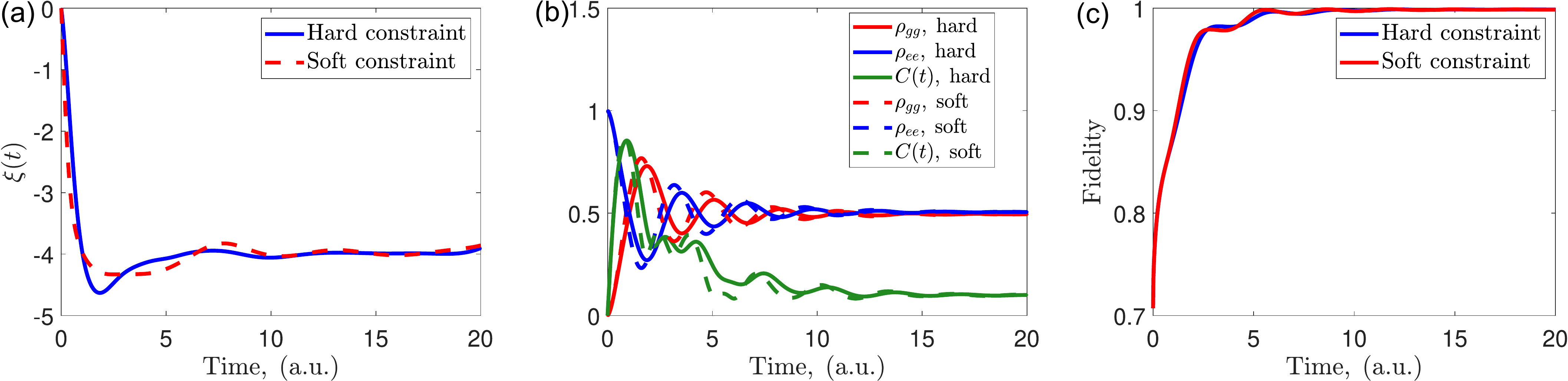}\\
	\caption{Comparison between soft ($\lambda = 1$) and hard constraints used for initial conditions in terms of: (a) control function $\xi(t)$, (b) dynamics, and (c) quantum fidelity. We used the same set of parameters presented in the main text for the simulations.}
	\label{figic}
\end{figure*}

\section{Background of physics-informed neural networks}

Physical laws are usually given in the form of differential equations. Neural Networks (NNs), on the other hand, provide closed-form solutions allowing the computation of analytical derivatives practically by using a technique called automatic differentiation \cite{autodiff_nips2017}. Having the NN's predictions and the derivatives in an analytical form, a  physical law can directly be encoded in the loss function, that is, when the loss function converges to zero, the differential equation is approximately satisfied. These machine learning models are usually called physics-based or physics-informed NN (PINNs).
For example,  driven dynamical systems are described by ordinary differential equations of the general form:
\begin{align}
  \label{eq:diffOperator}
D\left(t, {\bf x}, \dot {\bf x}, f(t) \right) = 0,
\end{align}
where $t$ is the independent time variable, and ${\bf x}(t)$ is a vector that denotes the dependent variables subjected to certain initial conditions ${\bf x}(0)= {\bf x}_0$. In Eq. (\ref{eq:diffOperator}), $D$ is an arbitrary function of $t$, ${\bf x}(t)$,   the first time derivative $\dot {\bf x}$, and of  the forcing time-dependent function $f(t)$. Similarly to the systems investigated in this study,  Eq. (\ref{eq:diffOperator}) describes a non-autonomous system.

As shown in Fig. 1 of the main manuscript, we considered that the solutions {\bf x} are parameterized by a NN where $t$ is the input to the network, and each output corresponds to a different variable of ${\bf x}$. Hence,  solving the differential equations of Eq. (\ref{eq:diffOperator}) is reduced to an optimization problem of the form:
\begin{align}
  \label{eq:loss0}
  \underset{\text{NN parameters}}{\arg\min} \left( \sum_{i=0}^{M}
  \Big( D\left(t_i, {\bf x}_i, \dot {\bf x}_i, f(t_i) \right)  \Big)^2 \right), 
\end{align}
where ${\bf x}_i = {\bf x}(t_i)$, and ${\dot {\bf x}}_i$ is analytically computed using automatic differentiation.  
The summation in Eq. (\ref{eq:loss0}) is over all the time points $M$. It defines a loss function which minimization yields NN parameters (weights and biases) that construct a neural solution ${\bf x}$ that approximately solves the differential equation (\ref{eq:diffOperator}). The initial conditions can be identically satisfied through a hard-constraint or approximately satisfied using a soft-constraint. In the main manuscript (in the text and in Fig. 1), it is explained how the initial conditions are identically satisfied through a specific parametrization. Later in Supplemental Material, we explain the soft-constraint approach and compare both in terms of the NN's performance. 
Subsequently, during the optimization (training) phase the NN is encouraged to make predictions that satisfy the underlying differential equations; thus, the predictions are approximate solutions to the physical problem. Although data are not required in this process, when they are available can be considered in an extra term in the loss function--- usually in the form of a mean square error  \cite{Raissi2019}.  An important advantage of PINNs over conventional numerical methods is that any known information (such as symmetries, part of the solution, conservation laws, etc) can potentially be embedded in the loss function as a soft-constraint, improving the training and the accuracy of NN's solutions \cite{Mattheakis2020}.

It has been shown that NNs can learn very compact and rich representations, making PINNs a powerful tool for learning complex solutions, like high-dimensional many-body wavefunctions \cite{Carleo2017,Cai2018,Choo2018, manybodyPRRes2023}. Moreover, PINNs scale well in high dimensions, suffering less than conventional numerical methods from the curse of dimensionality \cite{Sirignano2018, Zeng2022}. In the study, we introduce a novel physics-based architecture in the lines of PINNs. Our model can discover not only the solutions ${\bf x}$ of Eq. (\ref{eq:diffOperator}) but also the forcing function when a target final state is given. This target state plays the role of data and is satisfied through an extra loss function (Eq. (4) in the main manuscript). Additional constraints, like minimizing the amplitude of the external force (the case of this work that is discussed in the main manuscript), can be introduced as extra terms in the loss function. 

\section{Control and dynamical analysis in a two-level system}

For the two-level system, we use the following loss function

\begin{equation}
L = \lvert\lvert \dot{\mathbf{x}}-A(\lambda,\mathbf{u}(t))\mathbf{x} \rvert\rvert^2  + \lvert\lvert \mathbf{x} - \mathbf{x}_d \rvert\rvert^2  + L_{\rm reg},
\end{equation}
where $\mathbf{x}_d  = (1/2,1/2,0,0)^T$ is the desired target state vector which correspond to $\rho_d$. The first, second, and third terms on the right-hand side of the above equation imposes the Markovian master equation, the state preservation problem ($\rho(t) \rightarrow \rho_d$), and the constraint for off-diagonal elements ($\rho_{eg} \rightarrow 0$), respectively. From the equation of motion $\dot{\mathbf{x}} = A \mathbf{x}$ given in the section \textit{Two-level system of the main text}, we found:

\begin{eqnarray}
 \dot{x}_1 &=& \gamma_e - 2\Gamma x_1 -2 \omega_x x_4, \label{TLS1} \\
 \dot{x}_3 &=& -\Gamma x_3 - \left( 2\omega_z + \xi(t)\right)x_4, \label{TLS2}\\
 \dot{x}_4 &=& -\omega_x + 2\omega_x x_1 +  \left( 2\omega_z + \xi(t)\right)x_3 - \Gamma x_4, \label{TLS3}
\end{eqnarray}
where we have used $x_1+x_2 = 1$ ($\mbox{Tr}(\rho) = 1$) and $\Gamma = (\gamma_{\rm abs}+\gamma_{\rm em})/2$. First, we can analyze the steady state which implies that solutions satisfy $\dot{x}_i=0$ ($i=1,2,3$). 
In what follow, we demonstrate that the steady-state solution for the density matrix can be analytically solved in terms of the $\xi^{\rm ss} =\lim_{t \rightarrow \infty}\xi(t)$. Then, by numerically computing $\min_{\xi^{\rm ss}}(1-F(\rho(\xi),\sigma))$ ($F=1$ for a good control problem) we can compare the theoretical and neural network predictions for the steady-state. From numerical calculations using the PINN we found that $\xi(t)$ converges to a stable value, then we can asseverate that the steady state for $\xi(t)$ exist. Let us define $x_i^{\rm ss}$ and $\xi^{\rm ss}$ as the steady-state solutions, from equations~\eqref{TLS1}-\eqref{TLS3}, we obtain:

\begin{eqnarray}
\left( \begin{array}{ccc}
    2\Gamma & 0 & 2\omega_x \\
    0 & \Gamma  &  2\omega_z + \xi^{\rm ss}\\
    2\omega_x  &  2\omega_z + \xi^{\rm ss}& -\Gamma \\
\end{array}\right) \left(\begin{array}{c}
     x_1^{\rm ss}  \\
     x_3^{\rm ss}  \\
     x_4^{\rm ss}  
\end{array} \right) = \left( \begin{array}{c}
     \gamma_e  \\
     0  \\
     \omega_x  
\end{array}\right).
\end{eqnarray}

By solving the above linear system, we get the following analytical solutions for the steady-states in terms of $\xi^{\rm ss}$:

\begin{eqnarray}
 x_1^{\rm ss}&=& {\gamma_{\rm em} \over \gamma_{\rm em} + \gamma_{\rm abs}} + {4 w_x^2(\gamma_{\rm abs}-\gamma_{\rm em}) \over (\gamma_{\rm abs}+\gamma_{\rm em}\Delta(\xi^{\rm ss}))}, \\
 x_3^{\rm ss} &=& {4 w_x (\gamma_{\rm abs}+\gamma_{\rm em})(2w_z + \xi^{\rm ss}) \over (\gamma_{\rm abs}-\gamma_{\rm em}) \Delta(\xi^{\rm ss}) }, \\
 x_4^{\rm ss} &=& -{2w_x(\gamma_{\rm abs}-\gamma_{\rm em}) \over \Delta(\xi^{\rm ss})}, \\
 \Delta(\xi^{\rm ss}) &=& 4 \Gamma^2 + 8w_x^2 +16 w_z^2 + 16 w_z \xi^{\rm ss} + 4 (\xi^{\rm ss})^2.
\end{eqnarray}

For the particular case $w_x = 1$, $w_z = 2$, $\gamma_{\rm abs} = 0.1$, and $\gamma_{\rm em} = 0.3$ it follow that 

\begin{eqnarray}
 x_1^{\rm ss} &=& {3 \over 4} - {25 \over 2\Delta(\xi^{\rm ss})},  \\
 x_3^{\rm ss} &=& -{25(\xi^{\rm ss}+4) \over \Delta(\xi^{\rm ss})}, \\
 x_4^{\rm ss} &=& -{5 \over 2 \Delta(\xi^{\rm ss})}, \\ 
 \Delta(\xi^{\rm ss}) &=& 451 + 200 \xi^{\rm ss} + 25 (\xi^{\rm ss})^2.
\end{eqnarray}
Therefore, since the target state is defined as $\rho_d = (1/2)(|e\rangle\langle e|+|g\rangle \langle g|)$, or equivalently $x_d = (1/2, 1/2, 0, 0)^T$, we note that in order to obtain $x_3^{\rm ss} = 0$ it is necessary to have $\xi^{\rm ss} = -4$, as is shown in the inset of Figure~4 in the main text. In order to corroborate this previous observations rigorously,  we impose that the steady state must minimize the function $f(\xi) = 1-F(\rho(\xi),\sigma)$, where $\sigma=(1/2)(|e\rangle\langle e| + |g\rangle \langle g|)$ is the target state, $F(\rho(t),\sigma)=[\mbox{Tr}([\rho^{1/2}(t)\sigma \rho^{1/2}(t)]^{1/2})]^2$ is the quantum fidelity, and $\rho(\xi^{ss})$ is the steady state density matrix for $\xi^{\rm ss}$, and is given by 

\begin{equation}
    \rho(\xi^{\rm ss}) = \left(\begin{array}{cc}
       x_1^{\rm ss}   & x_3^{\rm ss}+ix_4^{\rm ss} \\
       x_3^{\rm ss}-ix_4^{\rm ss}   & x_2^{\rm ss} 
    \end{array} \right),
\end{equation}
where $x_2^{\rm ss}=1-x_1^{\rm ss}$. By employing a nonlinear optimization package of MATLAB we minimize $f(\xi) = 1-F(\rho(\xi),\sigma)$ over all possible values of $\xi^{\rm ss}$, and we obtain%By minimizing $\min_{\xi^{\rm ss}}(1-F(\rho(\xi),\sigma))$}, we obtain

\begin{equation}
\xi^{\rm ss} = -4, \quad \mbox{and} \quad  F(\rho(-4),\sigma) = 0.9988.
\end{equation}
By setting the value $\xi^{\rm ss} = -4$, we obtain the best performance for the steady state solution in terms of the quantum fidelity ($F\approx 1$), which implies that $x_{1}^{\rm ss} = 0.549$, $x_3^{\rm ss} = 0$, and $x_4^{\rm ss} = 0.049$.

\section{Initial conditions through hard and soft constraints}

\begin{figure}[ht]
	\centering
	\includegraphics[scale=0.6]{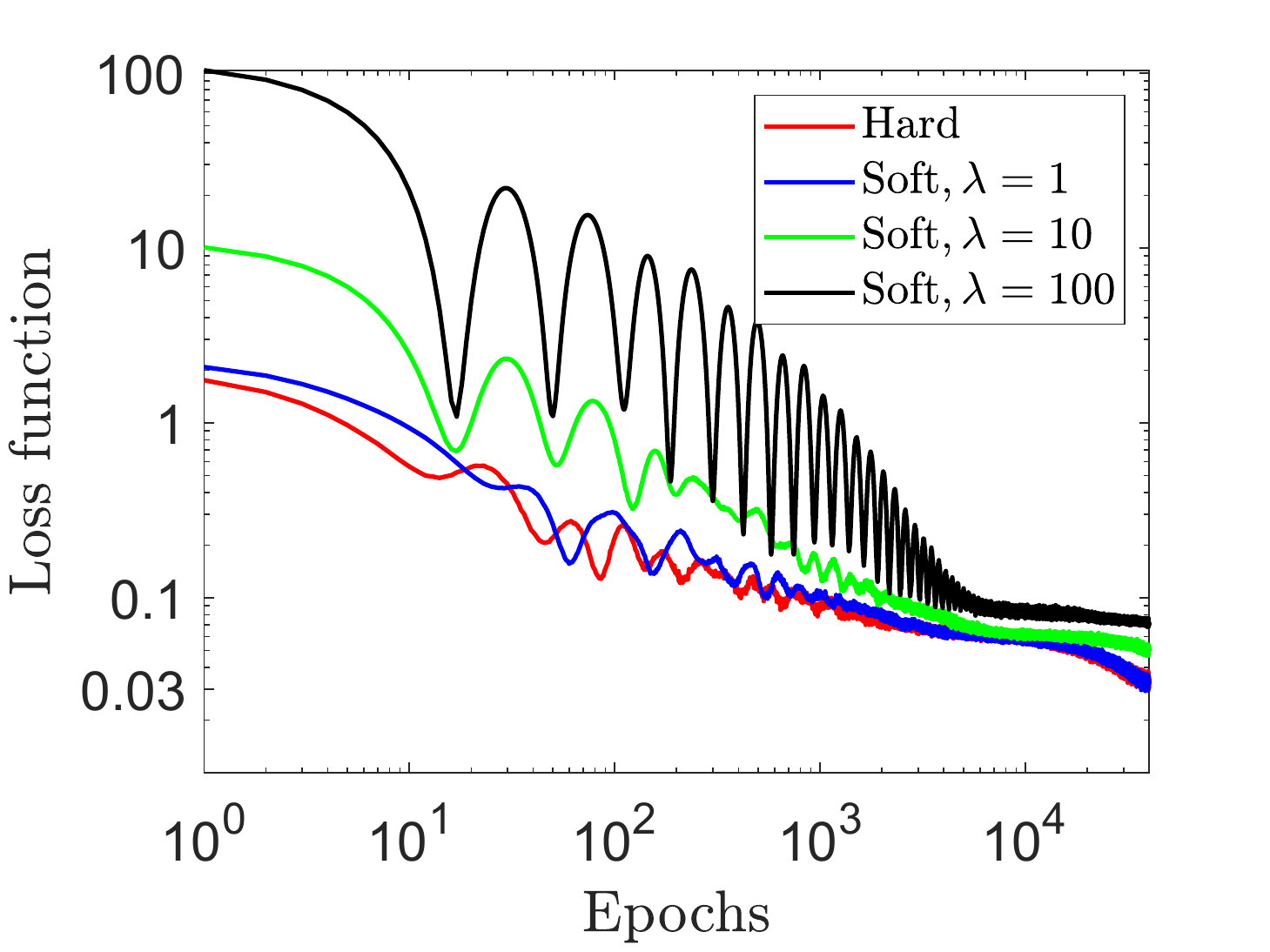}
	\caption{Comparison between soft and hard constraints in terms of the convergence of the loss function.}
	\label{fig:ComparisonLossFunction}
\end{figure}

In this section we compare two different approaches to incorporate initial conditions in our neural network architecture. In general terms, we have two possibilities to impose initial conditions, namely, i) use the parametric solution $\mathbf{x}(t) = \mathbf{x}(0)+f(t)\mathbf{N}_x(t)$ and $\mathbf{u}(t) = \mathbf{u}(0)+f(t)\mathbf{N}_u(t)$ with $f(0)=0$ (\textbf{hard constraint}) or ii) impose initial conditions into the loss function (\textbf{soft constraint}). To implement the soft constraint (without using the parametrization), we introduce the following loss function

\begin{equation}
    L_{\rm soft} = L_{\rm model} + L_{\rm control} + L_{\rm const} + L_{\rm reg} + L_{\rm ic} = L + \lambda L_{\rm ic},
\end{equation}

where $\lambda$ controls the soft constraint contribution, $L= L_{\rm model} + L_{\rm control} + L_{\rm const} + L_{\rm reg}$ is the same loss function presented in the main text, where we only added the term $L_{\rm ic}$ to take into account the initial conditions in the loss function. In the soft constraint approach, the term $L_{\rm ic}$ is defined as:

\begin{equation}
    L_{\rm ic} = \left[\lvert\lvert \mathbf{x}(t_1)-\mathbf{x}(0) \rvert\rvert^2 +   \lvert\lvert \mathbf{u}(t_1)-\mathbf{u}(0) \rvert\rvert^2 \right],
\end{equation}

where $\mathbf{x}(t)$ and $\mathbf{u}(t)$ are the vector and control functions, respectively, and $t_1$ is the initial time. To make a further comparison, we now solve the two-level system using these two different approaches (hard and soft). In Fig.~\ref{figic} we plot a comparison between the soft (for $\lambda = 1$) and hard constraints for the neural network solution of the two-level system presented in the main text. In Fig.~\ref{figic}(a), we observe that both control functions $\xi(t)$ are very similar leading a temporal behavior shown in Fig.~\ref{figic}(b). We note that both methods work equally well in terms of quantum fidelity, see Fig.~\ref{figic}(c). However, by changing the parameter $\lambda$ we observe that the convergence of the soft loss function decreases when $\lambda$ increases, as shown in Fig.~\ref{fig:ComparisonLossFunction}.

\section{Energy efficiency of the control scheme}
Let us consider a general Markovian open quantum system described by the master equation ($\hbar = 1$)
\begin{equation}
    \dot{\rho} = -i [H_0+H_c(t),\rho] + \mathcal{L}(\rho),
\end{equation}
where $H_0$ and $H_c(t)$ are the bare and control Hamiltonians, respectively. Here, $\mathcal{L}(\rho)$ describes the loss term in the Lindblad master equation. Then, the average energy of the system can be defined as $\langle E \rangle = \mbox{Tr}(H(t)\rho)$, with $H(t) = H_0 + H_c(t)$. The average power can be written as 
\begin{eqnarray}
    {d U \over dt} &=&  \mbox{Tr}(\dot{H}_c(t)\rho(t)) + \mbox{Tr}(H(t)\dot{\rho}) \nonumber \\
    &= &\dot{W}(t) + \dot{Q}. 
\end{eqnarray}

The term $\dot{W} = \mbox{Tr}(\dot{H}_c(t)\rho(t))$ is the power done by the system and $\dot{Q} = \mbox{Tr}(H(t)\dot{\rho}) $ is the rate of heat induced by the reservoir of the two-level system. Thus, by direct integration we obtain the expression for the work $W(t)$ and heat $Q(t)$:

\begin{eqnarray}
    W(t) &=& \int_{0}^{t} \mbox{Tr}(\dot{H}_c(t)\rho(t)) \,d \tau, \\ 
    Q(t) &=& \int_{0}^{t} \mbox{Tr}(H(t)\dot{\rho}) \,d \tau.
\end{eqnarray}

The time-dependent Hamiltonian competes with the heat flow induced by the environment. Therefore, we define $\eta = |W(t)/Q(t)|$ as the fraction of work done over the system compared to the internal heat that enters the system. \\

\begin{figure*}[ht]
\centering
\includegraphics[width=1 \linewidth]{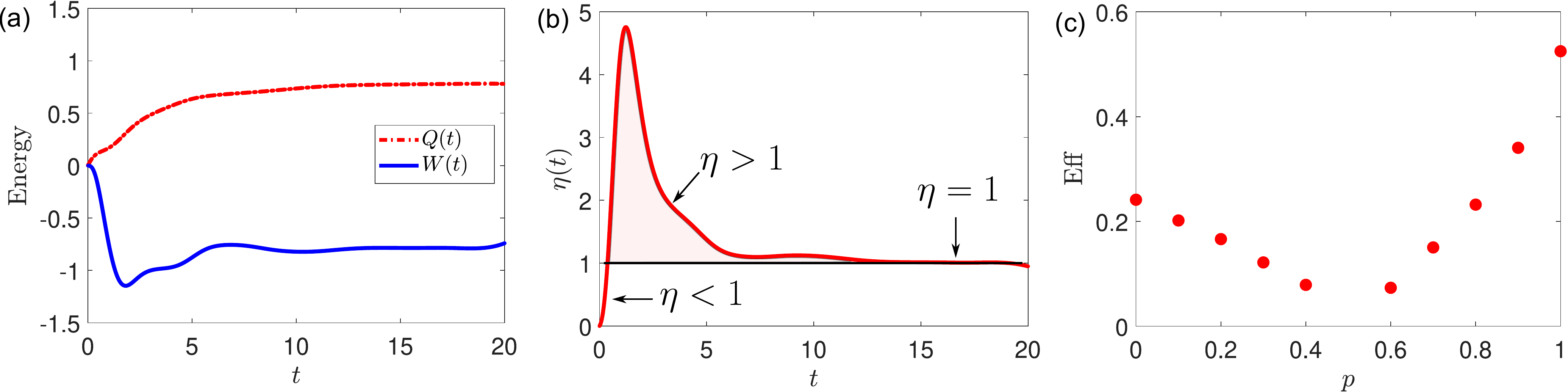}
\caption{(a) Heat and work as a function of time for the two-level system. (b) Parameter $\eta = |Q(t)/W(t)|$ as a function of time. (c) Efficiency parameter defined in Eq.~\eqref{Eff} for parametrized initial states $|\Psi(0)\rangle = p|g\rangle \langle g| + (1-p)|e\rangle \langle e|$, where $0 \leq p \leq 1$.}
\label{fig:FigureEnergy}
\end{figure*}

For the particular case of the two-level system presented in the main text, the work is given by 
\begin{equation}
    W(t) = \int_{0}^{t}\dot{\xi}(\tau) \rho_{ee}(\tau) \, d\tau,
\end{equation}
where $\xi(t)$ is the control field. In Figure~\ref{fig:FigureEnergy}(a) we plot the ratio the quantities $Q(t)$ and $W(t)$ for the two-level system. Also, in Figure~\ref{fig:FigureEnergy}(b) we show the temporal behavior of $\eta(t) = |W(t)/Q(t)|$ for the same parameters used in Figure.~3 of the main text. We note that we must apply a large quantity of work to overcome the effect of the internal heat. However, as time increases, the ratio $\eta$ converges to 1, showing that the applied work compensates for undesired heat effects. Consequently, the system reaches a stable and free-losses state, which is required to preserve the state over time. \\

Now, we shall elaborate on a measure to quantify the efficiency of the control protocol in terms of energy considerations. By analyzing Figure~\ref{fig:FigureEnergy}(b), we note that an efficient quantum control solution can be recognized as the one that minimizes the colored area for the curve $\eta-1$ for $\eta>1$. In other words, if we control a quantum system without inverting a large quantity of energy, then $W_{\rm app} \approx Q$, and therefore $\eta>1$ only in a short time domain. Let us define the integral, $I_1 = \int_{\eta >1}(\eta(\tau)-1)\; d\tau$ as a measure of the region where $\eta>1$. In the best scenario, $I_1 = 0$, implying that the energy efficiency will be equal to one. In addition, the total area of the curve $\eta$ will be represented by $I_2=\int \eta(\tau) \; d\tau$. Based on these observations, we define the energy efficiency over the whole process as 
$\rm{Eff} = 1-I_1/I_2$, or equivalently

\begin{equation} \label{Eff}
    \rm{Eff} = 1- {\displaystyle{\int_{\eta >1}(\eta(\tau)-1) \; d\tau }\over \displaystyle{\int \eta(\tau) \; d\tau}}, \quad 0 \leq \rm{Eff} \leq 1. 
\end{equation}

Since of $I_1<I_2$ (by construction) and $I_1=0$ is the minimum value, it follow that $0 \leq \rm{Ef} \leq 1$. Using this definition we have that $\rm{Eff}=0$ ($\rm{Eff}=1$) is the worst (best) scenario. For the two-level system, we obtain that $\mbox{Ef} = 0.5249$. Now, if we consider the initial state parametrization $\rho(0) = p|g\rangle\langle g| + (1-p)|e\rangle\langle e|$, we can calculate the energy efficiency in terms of the mixing parameter $0\leq p \leq 1$. In Figure~\ref{fig:FigureEnergy}(c), we plot the efficiency of the PINN for $p = \{0,0.1,0.2,0.3,0.4,0.5,0.6,0.7,0.8,0.9,1\}$. Note that the efficiency is not reported for $p=0.5$ since in such a case, the initial state is equal to the target state. 

\section{Control and dynamical analysis in a three-level system}

As it was stated in the main text, our PINN only handles real-valued functions. Therefore, we rewrite the master equation using the state vector $\vec{z}$ = $(\rho_{11}$,$\rho_{22}$,$\rho_{33}$,$\mbox{Re}[\rho_{12}]$,$\mbox{Im}[\rho_{12}]$,$\mbox{Re}[\rho_{13}]$,$\mbox{Im}[\rho_{13}]$,$\mbox{Re}[\rho_{23}]$,$\mbox{Im}[\rho_{23}])^T$. Hence, the dynamical equations reads,

\begin{align}\label{z_equations}
&0=\dot{z}_{1} + \Omega_p(t) z_7, \nonumber \\
&0=\dot{z}_{2} + \Omega_s(t) z_9, \nonumber \\
&0=\dot{z}_{3} - \Omega_p(t) z_7 - \Omega_s(t) z_9, \nonumber \\
&0=\dot{z}_{4} + \delta z_5 + (\gamma_1+\gamma_2)z_4 +{\Omega_p(t) \over 2}z_9- {\Omega_s(t) \over 2}z_7, \nonumber \\
&0=\dot{z}_{5}  - \delta z_4 + (\gamma_1+\gamma_2)z_5 + {\Omega_p(t) \over 2}z_8 - {\Omega_s(t) \over 2}z_6, \nonumber \\
&0=\dot{z}_{6} + \Delta_1 z_7 + (\gamma_1+\gamma_3)z_6 + {\Omega_s(t) \over 2}z_5, \nonumber \\
&0=\dot{z}_{7} - \Delta_1 z_6 + (\gamma_1+\gamma_3)z_7 +{\Omega_p(t) \over 2}z_3 - {\Omega_p(t) \over 2}z_1 \nonumber \\
& - {\Omega_s(t) \over 2}z_4, \nonumber \\
&0=\dot{z}_{8} -\delta z_9 + \Delta_1 z_9 + (\gamma_2+\gamma_3)z_8 - {\Omega_p(t) \over 2}z_5, \nonumber \\
&0=\dot{z}_{9} + \delta z_8 - \Delta_1 z_8 + (\gamma_2+\gamma_3)z_9 - {\Omega_p(t) \over 2}z_4 - {\Omega_s(t) \over 2}z_2 \nonumber \\
&+ {\Omega_s(t) \over 2}z_3.
\end{align}

We can define the loss function that accounts for the model as $L_{\rm model}=\sum_{i=1}^9 \Vert L_i\Vert^2$, where $L_i$ are given by the right-hand side in Eq.~\eqref{z_equations}. The control loss function comes from the target state, such that
\begin{equation}
L_{\rm control}= \eta\sum_{i=1}^{M}\Vert z_2(t_i)-1 \Vert^2. 
\end{equation}

\begin{figure}[ht]
	\centering
	\includegraphics[scale=0.6]{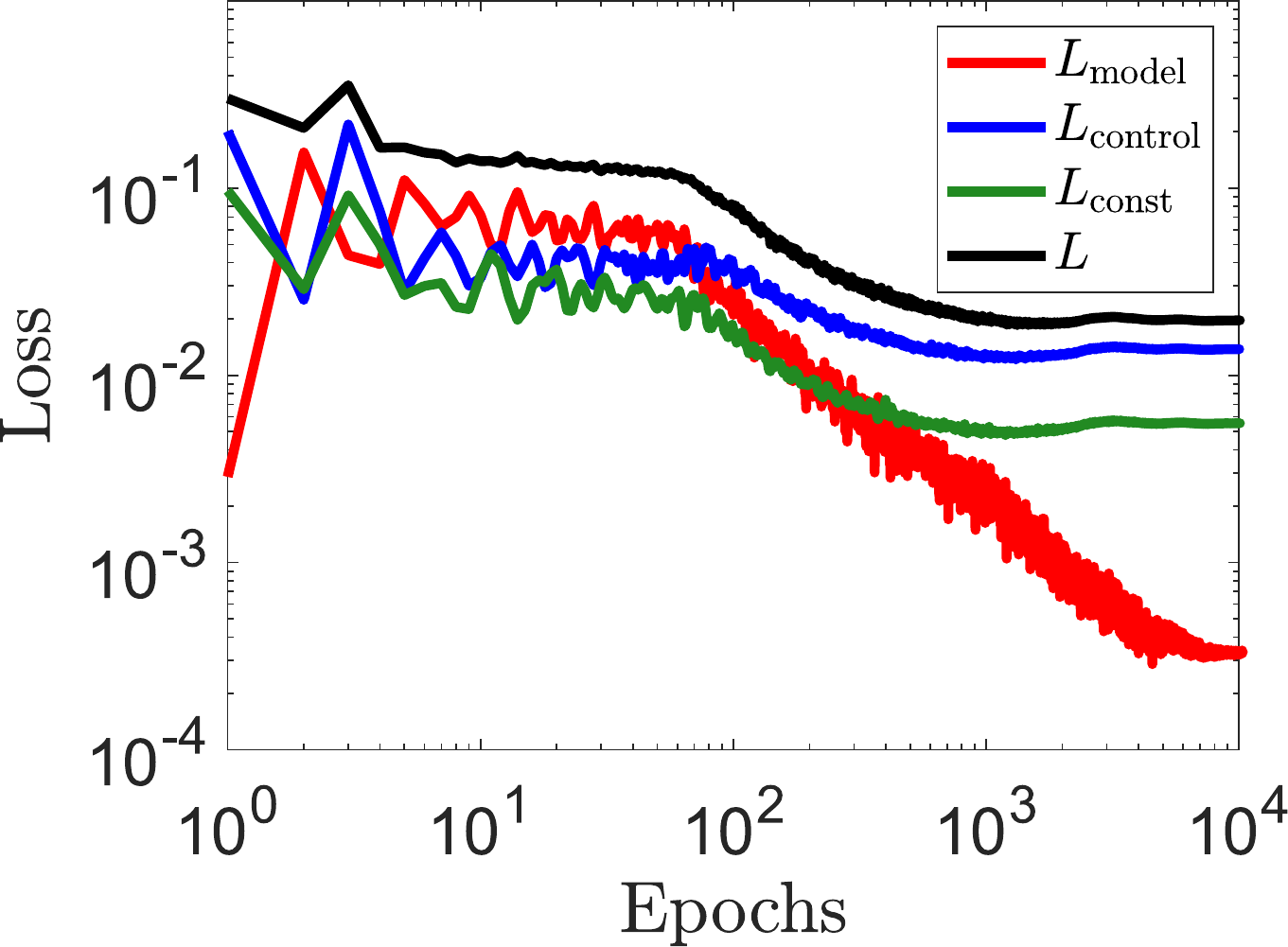}
	\caption{Convergence of different components of the loss function for the three-level system defined in Eq.~\eqref{LossThreeLevel}.}
	\label{fig:Loss}
\end{figure}

It is well-known that the master equation in the Lindblad form ensures that the density matrix will be Hermitian, positive, semi-definite, and $\mbox{Tr}[\rho]=1$. However, we found that enforcing the last constraint helps the PINN to find the best solution. Therefore, we add the following term to the loss function

\begin{equation}\label{L_const}
L_{\rm const}=\eta_c \sum_{i=1}^{M}(\Vert z_1(t_i)\Vert^2 +  \Vert z_3(t_i)\Vert^2),   
\end{equation}
which together with $L_{\rm control}$ enforces $\mbox{Tr}[\rho]=1$. Finally, we added a $l_2$-norm regularization $L_{\rm reg}=\chi\sum_{i}\Omega_i^2$ that penalizes the control fields being too large and delivers smooth functions. Henceforth, the PINN tries to minimize the overall loss function:
\begin{equation}
L= L_{\rm model} + L_{\rm control} + L_{\rm const} + L_{\rm reg}, \label{LossThreeLevel}
\end{equation}
with a learning rate that amounts to $8\times10^{-3}$. The weights in $L_{\rm control}$, $L_{\rm const}$ and $L_{\rm reg}$ ($\eta=0.2$, $\eta_c=0.1$ and $\chi=2.8\times 10^{-3}$, respectively) regulate the relevance of the control, constraint, and smoothness conditions comparing to the leading loss component $L_{\rm model}$. These hyperparameters were adjusted manually. In Figure~\ref{fig:Loss} we show the Loss function of the training process corresponding to Figure 4-(a) in the main text, where we also show the convergence of $L_{\rm model}$, $L_{\rm control}$ and $L_{\rm const}$.

We now focus on the transfer time. We numerically observe that the PINN reduces the time for having a successful population transfer as compared to the other sequences, see Table I in the main text. This particular feature of the PINN naturally shows up when enforcing the control and constraining loss functions upon each time point ($t_i$). In other words, the PINN itself minimizes the transfer time. To realize this, we compute the PINN with only the control and constraint loss functions enforced by the last thirty points. We found that the population transfer slows down, and the probability slightly decreases. Nevertheless, the PINN can still handle this scenario if we randomly sample the time grid for each epoch and then pick thirty points (or even ten) to enforce the control and constrain loss functions. We remark that this is only possible because PINNs do not require a structured mesh; thus, $t_i$ can be arbitrarily discretized.

%%%%%%%%%%%%%%%%%%%%%%%%%%%%%%%%%

\section{Pulse sequences for controlling a $\Lambda$-system.}

\begin{figure*}[ht]
\includegraphics[scale=0.2]{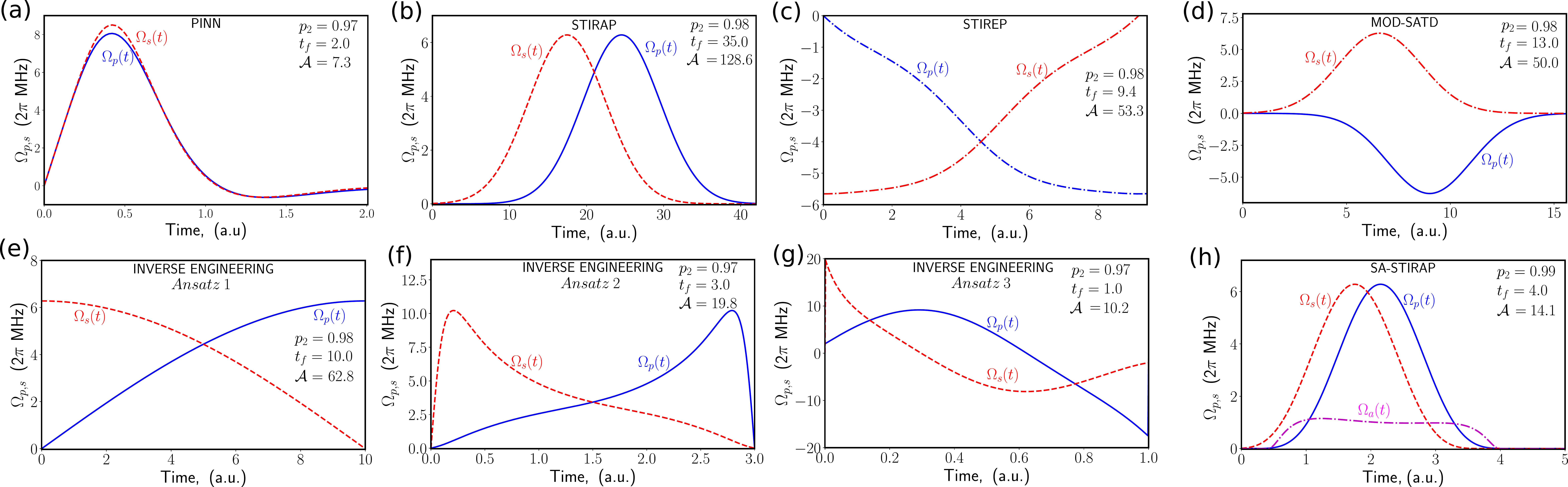}
\caption{Comparing all methods for population transfer from state $\ket{1}$ to $\ket{2}$.} 
\label{fig_protocols}
\end{figure*}

In this section, we detail some control protocols that exploit adiabaticity, shortcuts to adiabaticity, or inverse engineering. We aim to efficiently transfer population from an initially polarized state ($|1\rangle$) to a target state ($|2\rangle$) via an intermediary state ($|3\rangle$). To achieve this state-to-state transfer, two radiation fields are used to induce transitions $\ket{1}\leftrightarrow \ket{3}$ and $\ket{2}\leftrightarrow \ket{3}$, as shown in Fig.~2(b). In what follows, we provide a detailed derivation of each of these methods--- some of them already addressed by us in Ref.~\cite{Gonzalez} and included here to make the present paper self-contained. We remark that the optimization can be improved by increasing the control fields for most of these methods and our PINN. However, we restrict them to $\Omega_{s,p}\leq 10$ to fairly compare the methods. In Fig.~\ref{fig_protocols} we compare all methods for the population transfer between the states $\ket{1}$ and $\ket{2}$.

\subsubsection{Stimulated Raman Adiabatic Passage (STIRAP)}

First, we consider a well-known protocol for adiabatic transfer dubbed Stimulated Raman Adiabatic Passage (STIRAP)~\cite{Kuklinski1989,Bergmann1998,Vitanov2017}. Under the rotating wave approximation, the Hamiltonian for this system is given by,
\begin{equation}\label{H_stirap}
 H(t) = \delta \sigma_{22}+ \Delta_1 \sigma_{33}
 +\left(\frac{\Omega_{p}(t)}{2}\sigma_{31}+ \frac{\Omega_{s}(t)}{2}\sigma_{32}+ h.c. \right).
\end{equation}

Hereafter, we set to zero the two-photon detuning $\delta=0$ ($\Delta_1 = \Delta_2 = \Delta$). The eigenstates of  the Hamiltonian are:
\begin{align}
&\ket{\Phi_{+}}=\sin\theta\sin\phi\ket{1} + \cos\phi\ket{3} + \cos\theta\sin\phi\ket{2},\nonumber\\
&\ket{\Phi_{-}}=\sin\theta\cos\phi\ket{1} - \sin\phi\ket{3} + \cos\theta\cos\phi\ket{2}, \nonumber\\
&\ket{\Phi_{d}}=\cos\theta\ket{1} - \sin\theta\ket{2}, \label{Eigenstates}
\end{align}
 with instantaneous eigenvalues $E_d=0$ and $E_{\pm}(t)=\Delta/2 \pm \left(\Delta^{2}+\Omega_p^2(t)+\Omega_s^2(t)\right)^{1/2}/2$. The mixing angles are defined through the relations
\begin{equation}\label{theta}
\tan\theta(t)=\frac{\Omega_p(t)}{\Omega_s(t)}, \hspace{0.2cm} \tan 2\phi(t) = \frac{\sqrt{\Omega_p^2(t)+\Omega_s^2(t)}}{\Delta}.
\end{equation}

We note that the dark state ($\ket{\Phi_{d}}$) has no contribution from the excited state $\ket{3}$, so the population transfer from state $\ket{1}$ to state $\ket{2}$ is driven by the variation of the mixing angle $\theta(t)$. The latter implies that the Rabi frequencies $\Omega_p(t)$ and $\Omega_s(t)$ must be correlated. First, we note that $\ket{\Phi_{d}}$ coincides with $\ket{1}$ when $\theta(t)=0$, which is obtained from $\Omega_{p}(t)/\Omega_{s}(t)\longrightarrow 0$. Second, the population transfer is completed when $\ket{\Phi_{d}}$ coincides with $\ket{2}$ ($\theta(t)=\pi/2$), that is obtained from $\Omega_{p}(t)/\Omega_{s}(t)\longrightarrow \infty$. Hence, the population transfer is attained with a counterintuitive pulse order, i.e. $\Omega_s(t)$ precedes $\Omega_p(t)$.

As mentioned above, the STIRAP protocol follows an adiabatic evolution, usually slower than superadiabatic control protocols. Therefore, we now focus on three different modifications to STIRAP that use superadiabatic corrections~\cite{guery2019,giannelli2014,chen2010,sun2020}, inverse engineering~\cite{chen2012,Lai1996} and counteradiabatic approach~\cite{Baksic2016}.

\subsubsection{Inverse  Engineering}

Inverse engineering is one of the protocols that achieves high fidelity in shorter times. The main goal is to design the optimal control pulses $\Omega_{p}(t)$ and $\Omega_{s}(t)$. To determine these control fields, an invariant operator $I(t)$ is used, which satisfies the equation~\cite{chen2012,Lai1996},
\begin{equation}
\frac{\partial I(t)}{\partial t}+\frac{1}{i\hbar}[I(t),H_{0}(t)]=0.
\end{equation}

The condition $\Delta_1=0$ yields the following operator ($\hbar=1$),
\begin{align}
I(t)=& \frac{\Omega_{0}}{2}\begin{pmatrix}
   0 &  \varXi(t)    & \varUpsilon(t) \\
  \varXi(t) &   0        &  \varXi(t)  \\
 \varUpsilon^{\ast}(t) & \varXi(t)     & 0 \\
  \end{pmatrix},\label{IMatrix}
\end{align}
with $\varXi(t)= \cos\gamma(t)\sin\beta(t) $ and $\varUpsilon(t)=-i\sin\gamma(t)$. The time-dependent auxiliary parameters $\gamma(t)$ and $\beta(t)$ satisfy the following equations~\cite{chen2011},
\begin{align}
 \frac{d\gamma(t)}{dt}&=\frac{1}{2}(\Omega_{p}(t)\cos\beta(t)-\Omega_{s}(t)\sin\beta(t)),\nonumber\\
 \frac{d\beta(t)}{dt}&=\frac{1}{2}\tan\gamma(t)(\Omega_{s}(t)\cos\beta(t)+\Omega_{p}(t)\sin\beta(t)).
\end{align}

From the above equations one can find the optimal control fields as
\begin{align}
 \Omega_{s}(t)&=2(\frac{d\beta(t)}{dt}\cot\gamma(t)\cos\beta(t)-\frac{d\gamma(t)}{dt}\sin\beta(t)),\nonumber\\
 \Omega_{p}(t)&=2(\frac{d\beta(t)}{dt}\cot\gamma(t)\sin\beta(t)+\frac{d\gamma(t)}{dt}\cos\beta(t)).
\end{align}

Henceforward, we shall consider different Ansatzs for the auxiliary parameters $\gamma(t)$ and $\beta(t)$.\\

\textit{Ansatz 1:} For simplicity, we choose $\gamma(t)=\epsilon$ and $\beta(t)=\pi t/2t_{f}$ \cite{chen2012}, that leads to
\begin{align}
 \Omega_{s}(t)&=\frac{\pi}{t_{f}}\cot\epsilon\cos\left(\frac{\pi t}{2t_{f}}\right),\nonumber\\
 \Omega_{p}(t)&=\frac{\pi}{t_{f}}\cot\epsilon\sin\left(\frac{\pi t}{2t_{f}}\right).
\end{align}

For the numerical calculations we set $\epsilon=0.05$ and $t_f=10$. \\

\textit{Ansatz 2:} We choose the polynomial solution introduced in Ref.~\cite{chen2012},
\begin{equation}
  \beta(t)=\sum_{j=0}^{3}b_{j}t^{j},\quad \quad \gamma(t)=\sum_{j=0}^{4}a_{j}t^{j},
\end{equation}
where the coefficients $b_j$ and $a_j$ are determined from the initial conditions
\begin{align}
  \gamma(0)&=\epsilon, \quad \dot{\gamma}(0)=0,\quad \gamma(t_{f})=\epsilon, \quad \dot{\gamma}(t_{f})=0, \quad \gamma(t_{f}/2)=\delta,\nonumber\\
  \beta(0)&=0,\quad \dot{\beta}(0)=0, \quad\beta(t_{f})=\frac{\pi}{2},\quad \dot{\beta}(t_{f})=0.
\end{align}

To obtain an optimal and fast population transfer we set $\epsilon=0.02$, $\delta = \pi/10$ and $t_{f}=3$. We show our results for this ansatz in Table I in the main text. \\

\textit{Ansatz 3:} We choose the solution proposed in Ref.~\cite{Kiely2014}, that achieves successful population transfer in the presence of unwanted transitions. The parameters are,
\begin{eqnarray}
  \gamma(t)&=&-\frac{8(\pi-2d_{0})}{T^{4}}t^{4}+\frac{2(7\pi-16d_{0}+T)}{T^{3}}t^{3} \nonumber \\
  &&-\frac{5\pi-16d_{0}+3T}{T^{2}}t^{2}+t, \\
  \beta(t)&=& -\pi\left(\frac{t}{T}\right)^{3}+\frac{1}{2}(2\pi d_{1}+3\pi)\left(\frac{t}{T}\right)^{2} \nonumber \\
  && +\left(-\frac{1}{2}(2\pi d_{1}+3\pi+\frac{3\pi}{2})\right)\frac{t}{T}+d_{1}\sin\left(\frac{\pi t}{T}\right),
\end{eqnarray}
where $d_0=1.8$, $d_{1}=0.1$ and $T=1$. While this ansatz works fine, we did not include it in Table I due to the large control field, which we could not decrease below $10$ with a successful transfer rate. 

\subsubsection{Stimulated Raman Exact Passage (STIREP)}

Recently, a new protocol combines inverse engineering and optimization methods~\cite{Laforgue2022}. Here, the control pulses are determined by trajectories $\tilde{\phi}$ and read,
\begin{align}
 \tilde{\Omega}_{s}&= \Omega_{s}(t)/2\dot{\eta}= \cos\tilde{\phi}\sin\tilde{\theta}-\dot{\tilde{\phi}}\cos\tilde{\theta},\nonumber\\
 \tilde{\Omega}_{p}&=\Omega_{p}(t)/2\dot{\eta}=-\cos\tilde{\phi}\cos\tilde{\theta}-\dot{\tilde{\phi}}\sin\tilde{\theta},
\end{align}
with $\tilde{\phi}=\phi[\eta(t)]$. The trajectory $\tilde{\phi}$ is obtained by applying robust inverse optimization (RIO)~\cite{Laforgue2022}, which optimizes a cost function represented by the total area of the two pulses and is given by $\mathcal{A}_{t}=2\int_{\eta_{i}}^{\eta_{f}} |\dot{\eta}|\sqrt{\dot{\tilde{\phi}}^{2}+\cos^{2}\tilde{\phi}} \,d\eta $. By using the Euler-Lagrange equations and the Lagrange Multipliers method we obtain the following differential equations for the trajectory,,
\begin{eqnarray}
  \dot{y}_{1}&=&\dot{\tilde{\phi}}^{\pm}=y_{2},\nonumber\\
  \dot{y}_{2}&=&\ddot{\tilde{\phi}}^{\pm}=-(2y_{2}^{2}+\cos^{2}\tilde{\phi}^{\pm})\tan\tilde{\phi}^{\pm} \nonumber \\
  && \pm(\lambda_{0}\sec\tilde{\phi}^{\pm}\lambda_{1}\sin\eta-\lambda_{2}\cos\eta)\left(y_{2}^{2}+\cos^{2}\tilde{\phi}^{\pm}\right)^{3/2} = 0. \nonumber \\
\end{eqnarray}

For the initial condition $\dot{\tilde{\phi}}=0$ we end up with the Lagrange Multipliers $\lambda_0=0.394$, $\lambda_{1}=-0.064$ and $\lambda_{2}=0.283$.

\subsubsection{Modified Superadiabatic Transitionless Driving (MOD-SATD)}

The MOD-SATD~\cite{Baksic2016} protocol is a different alternative, which bypasses the adiabatic condition while counteracting the effect of the loss of adiabaticity. The fields in this protocol are parameterized according to~\cite{Baksic2016}
\begin{align}\label{mod-satd}
 \Omega_{p}(t)&=-\Omega^\prime(t)\sin\theta^\prime(t),\nonumber\\
 \Omega_{s}(t)&=\Omega^\prime(t)\cos\theta^\prime(t),
\end{align}
where $\theta^\prime(t)=\theta(t)-\arctan[g_{x}(t)/(\Omega(t)+g_{z}(t))]$, $\Omega^\prime(t)=\sqrt{[\Omega(t)+g_{z}(t)]^{2}+g_{x}^{2}(t)}$, $\mu(t)=-\arctan[\dot{\theta}/(\Omega(t)+g(t)/\sigma_m)]$, $g_{x}(t)=\dot{\mu}$, $g_{z}(t)=-\Omega(t)-\dot{\theta}/\tan(\mu)$, $g(t)=A/\cosh(\zeta t)$ with  $A=1/40$, $\zeta=9/(10\sigma_m)$ and $\sigma_m=2.0$~$\mu$s. For the Gaussian fields considered in STIRAP one finds that~\cite{Baksic2016}
\begin{eqnarray}
 \theta(t)&=&\arctan\left[\exp\left(t_{d}t/\sigma^{2}\right)\right],\nonumber\\
 \Omega(t)&=&\Omega_{0}\exp\left(-\frac{t^2+t_{d}^{2}/4}{2\sigma^{2}}\right)\sqrt{2\cosh\left(\frac{t_{d}t}{\sigma^{2}}\right)},
\end{eqnarray}
with $t_{d}=6/5\sigma$ and $\Omega_0/2\pi=1$~MHz.

\subsubsection{Superadiabatic STIRAP (SA-STIRAP)}

For completeness, we also show the SA-STIRAP sequence. However, we do not consider it in the main text (Table I) because it relies on a pulse connecting states $\ket{1}$ and $\ket{2}$, which we have not allowed for our PINN.

The superadiabatic approximation requires an external control Hamiltonian $H_{\rm c}(t)$~\cite{giannelli2014,chen2010,sun2020}, such that
\begin{equation}
 H_{\rm sa}(t) = H(t) + H_{c}(t),
\end{equation}
where $H(t)$ is the original Hamiltonian in Eq.~\eqref{H_stirap}. The superadiabatic correction reads ($\hbar=1$),
\begin{eqnarray}\label{H11}
 H_{\rm c}(t) &=& i\sum_{n=\pm,d}\left[ \ket{\partial_{t}\Phi_{n}(t)}\bra{\Phi_{n}(t)} \right. \nonumber \\
 && - \left. \bra{\Phi_{n}(t)}\partial_{t}\Phi_{n}(t)\rangle \ket{\Phi_{n}(t)}\bra{\Phi_{n}(t)}\right],
\end{eqnarray}
where $\ket{\Phi_{n}}$ are the eigenstates $H(t)$, as given in Eq.~\eqref{Eigenstates}. After straightforward calculations we obtain the superadiabatic correction 
\begin{equation}
 H_{\rm c}(t) =\frac{1}{2}\left(i\Omega_{a}(t)\sigma_{21} -i\Omega_a^*(t)\sigma_{12}\right),
\end{equation}
with $\Omega_{a}(t)\equiv2d\theta(t)/dt$, and $\theta(t)$ given in Eq.~\eqref{theta}. Instead of Gaussian fields (as we used in the STIRAP protocol), here we use $\Omega_p(t)=\Omega_{0}\sin^{4}(\pi(t-\tau)/T)$ and $\Omega_s(t)=\Omega_{0}\sin^{4}(\pi t/T)$, with $\Omega_0/2\pi=1$, $\tau=0.1T$ and $T=4$.

%%%%%%%%%%%%%%%%%%%%%%%%%%%%%%%%%%%%%

\section{Optimization with two-photon detuning}

\begin{figure}[ht]
\includegraphics[scale=0.5]{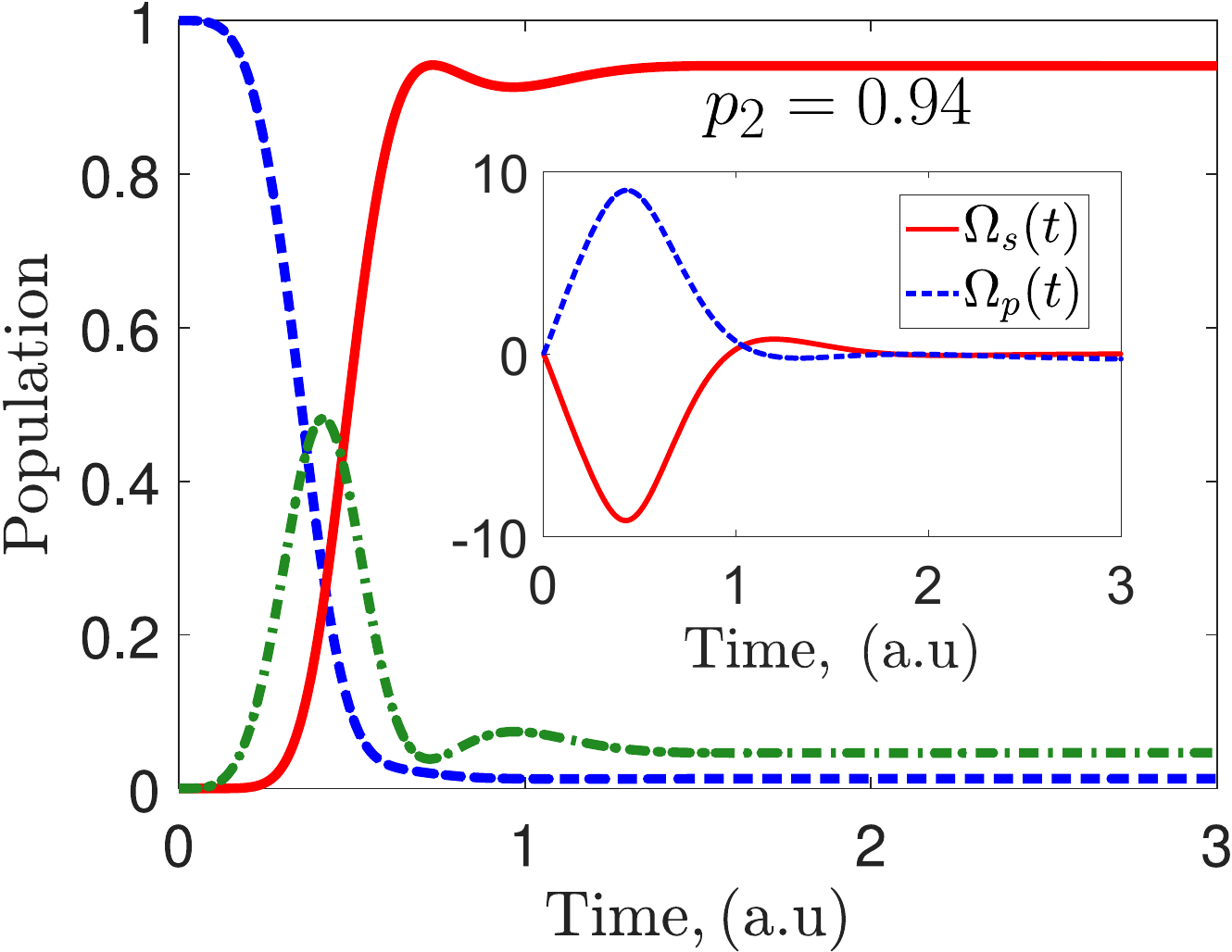}
\caption{Training the PINN considering one- ($\Delta_1/2\pi=0.2$) and two-photon ($\delta/2\pi=0.2$) detuning can easily improve the population transfer from $p_2=0.93$ to $p_2=0.94$.} 
\label{fig:detun}
\end{figure}

The one-photon detuning does not have a detrimental role in the transfer protocol. For instance, it is known that it does not affect STIRAP in the adiabatic limit~\cite{Vitanov2017} (We performed numerical simulations and observed no difference with the one-photon detuning. ). In contrast, the two-photon detuning ($\delta$) hurts the success in the population transfer~\cite{Romanenko1997,Vitanov2017}. Back to STIRAP, neither the dark state nor the null eigenvalue are available, forcing variations in the sequence (or a total departure from it) to achieve a successful transfer. Our PINN, on the other hand, can be easily trained to counteract this effect. It only requires setting the new value of $\delta$ before the training process, and then the PINN automatically updates the weights in the optimization and delivers a population transfer with a final probability $p_2=0.94$, see Fig.~\ref{fig:detun}. Furthermore, it is worth noticing that the pulse sequence found by the PINN does not experience a significant change. This means the sequence is robust without a subsequent optimization ($p_2=0.93$ in the main text). It is important to note that along the manuscript with have fixed the regularization parameter $\chi=2.8\times 10^{-3}$. However, decreasing it down to $\chi=1\times 10^{-3}$ will increase the control fields and deliver a population transfer with $p_2=0.97$.

%%%%%%%%%%%%%%%%%%%%%%%%%%%%%%%%%%%%%

\section{Four-level system}

The optimization of control pulses is, in general, system-dependent. This means that once we have the optimal pulse sequence for a $\Lambda$-system, the extension to a four-level system is not straightforward. For instance, the new system involves more non-linear equations and the possibility of cross-talk to the newly added state. Henceforth, any analytical approach for optimization becomes a hard task. In this section we show that our PINN can handle a four-level system without modification to the architecture of the network. Then, we feed the PINN with the new set of differential equations, and we add a new constraint in $L_{\rm const}$~\eqref{L_const} for the population in the fourth state $\Vert z_4\Vert^2$. Our goal is to train a PINN that succeeds in transferring population from state $|1\rangle$ to state $|2\rangle$, with, (\textit{i}) minimizing the population in the lossy intermediary state ($| 3 \rangle$), (\textit{ii}) minimizing the pulse area, and (\textit{iii}) minimizing cross-talk to state $|4\rangle$. Note that the last condition guarantees selectivity and it was not required for the three-level system.    

\begin{figure}[ht]
\includegraphics[scale=0.35]{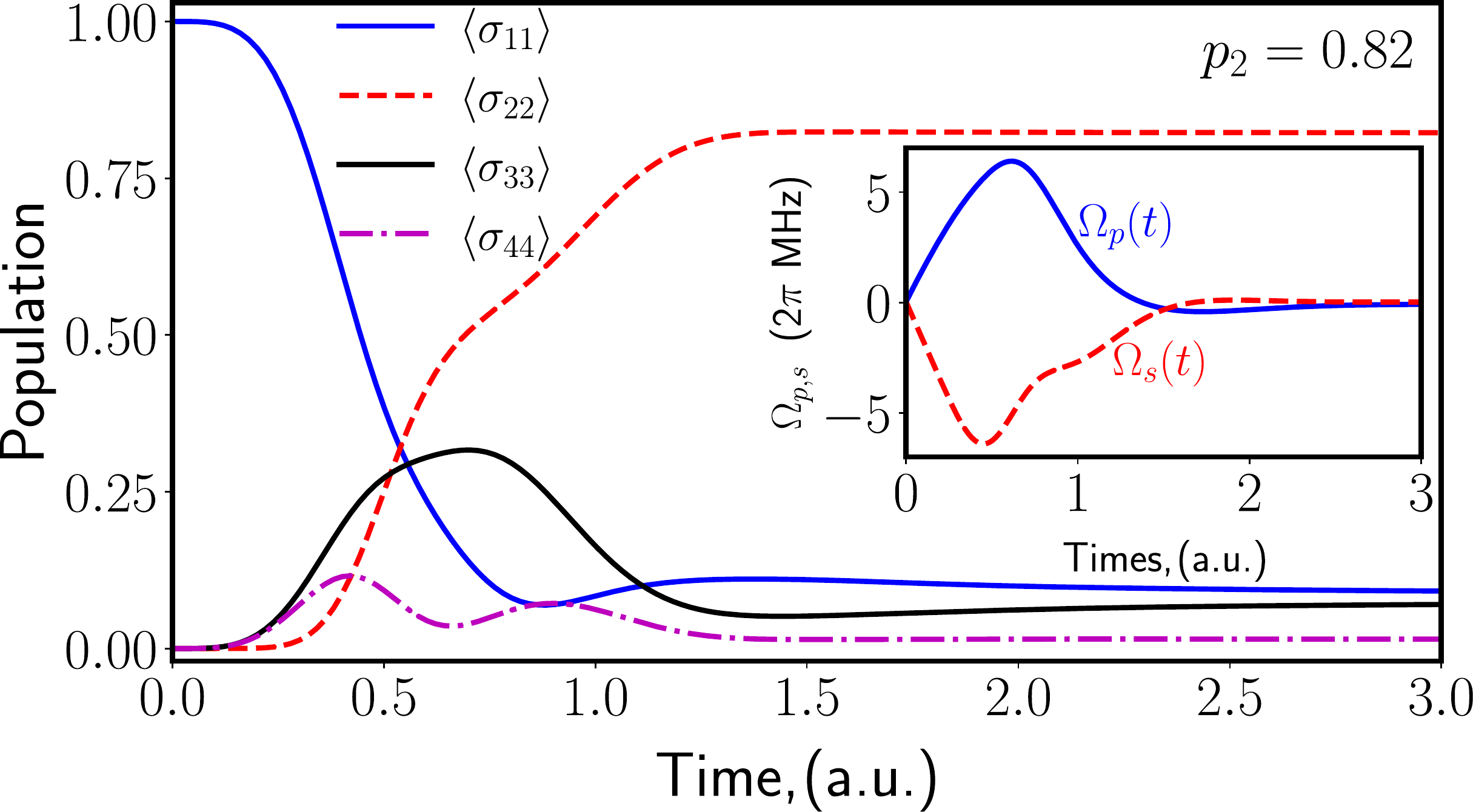}
\caption{PINN delivers a successful population transfer method that can be scaled to a four-level system without changing the network's architecture. We set $\Delta_2=\Delta_3=\delta=0$ and $\Delta_4=6.79$.} 
\label{fig:4level}
\end{figure}

To begin with, we consider the Hamiltonian of the system in a multi-rotating frame and in the eigenstate basis ($\hbar=1$),
 \begin{eqnarray}
\hat{H} &=& \delta\sigma_{22} + \Delta_3\sigma_{33} + \Delta_4\sigma_{44} +\frac{\Omega_p(t)}{2}\left(\sigma_{13}+\sigma_{14}\right) \nonumber \\
&& +\frac{\Omega_s(t)}{2}\left( \sigma_{23} - \sigma_{24} \right) + h.c.,
 \end{eqnarray}
where $\Omega_p(t)$ and $\Omega_s(t)$ are the Rabi frequencies of the control fields and $\sigma_{ij}$ the ladder operators. The one-photon detunings are $\Delta_2=(E_3-E_2-\omega_s)$, $\Delta_3=(E_3-E_1-\omega_p)$, $\Delta_4=(E_4-E_1-\omega_p)$ and the two-photon detuning is $\delta_2=(\Delta_3-\Delta_2)$. $E_i$ are the eigenenergies and $\omega_p$ and $\omega_s$ are the frequency of the control fields. This Hamiltonian can be obtained from the interaction of a Nitrogen-Vacancy center with a Carbon-13 nuclear spin \cite{Gonzalez,Coto17}. In Fig.~\ref{fig:4level} we show the results for our population transfer method.

\begin{figure*}[ht]
	\centering
	\includegraphics[scale=0.5]{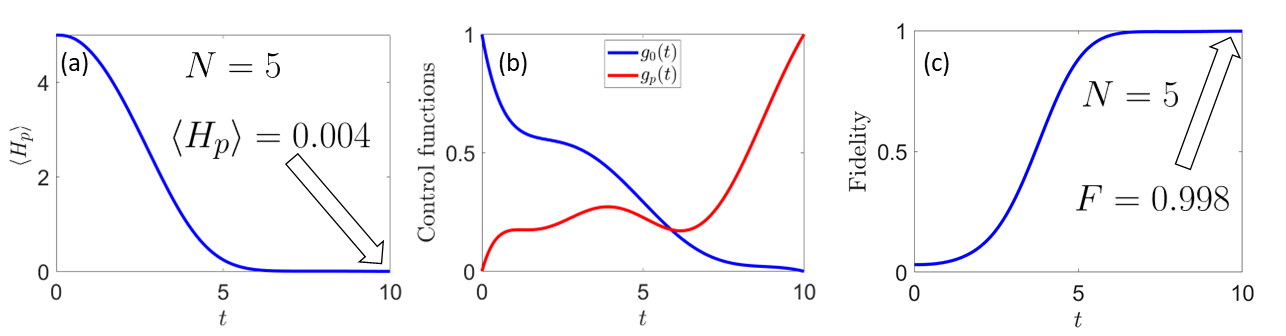}\\
	\caption{(a) Expectation value $\langle H_p \rangle$ as a function of time. (b) Control functions obtained from our PINN approach. (c) Fidelity of the protocol.}
	\label{Figure1}
\end{figure*}

\section{Large quantum systems}

\subsection{Non-interacting qubits}

Let us consider the following Hamiltonians

\begin{equation}
H_0=-\frac{\omega_x}{2}\sum_{j=1}^{N}\sigma_{x,j}, \quad H_p= \frac{\omega_z}{2}\sum_{j=0}^{N}(1-\sigma_{z,j}). 
\end{equation}

where $\sigma_{x,j}$ and $\sigma_{z,j}$ are the Pauli matrices for $S=1/2$ of the $j$th qubit. Here, $N$ is the number of qubits, and $2^N$ is the dimension of the Hilbert space. All qubits are identical with $\omega_x, \omega_z >0$. Now, the target of this control problem is to minimize the expectation value of the final Hamiltonian. For that reason, we define

\begin{equation}
    L_{\rm control} = \langle H_p \rangle = \langle \Psi(t) | H_p | \Psi(t)\rangle,
\end{equation}

Let us consider that $u_1(t) = g_0(t)$ and $u_2 = g_p(t)$ are two control functions that defines a real control vector $\mathbf{u}(t) = (u_1(t),u_2(t))^T$. The whole system satisfies the Schr\"{o}dinger equation $i\hbar \partial_t |\Psi(t)\rangle = H(t)|\Psi(t)\rangle$, where $\ket{\Psi(t)}$ is the wavefunction. By using the decomposition into real and imaginary parts, \textit{i.e.}, $\ket{\Psi(t)} = \ket{\Psi(t)}^R+i\ket{\Psi(t)}^I$ and $H(t) = H^R(t)+iH^I(t)$ ($H= H^{\dagger}$), we get the following non-autonomous dynamical system ($\hbar = 1$)

\begin{equation}
\dot{\mathbf{x}} = A(t) \mathbf{x}(t), \quad A(t) = \left(\begin{array}{cc}
H^I(t) & H^R(t) \\
-H^R(t) & H^I(t)
\end{array} \right),
\end{equation}

where $\mathbf{x}(t) = (\ket{\Psi(t)}^R, \ket{\Psi(t)}^{I})^T \in \mathds{R}^{2^{N+1}}$ is the real state vector required for the artificial neural network, and $A(t) $ is the dynamical matrix that depends on the control vector. Thus, we can apply our methodology to the dynamical system and find optimal control functions to reach the ground state of the final Hamiltonian.\\

For illustration, let's consider the case of five qubits ($N=5$), where the dynamical system is highly non-trivial since we have 64 coupled equations of motion for the real vector $\mathbf{x}(t)$ ($2^{5+1} = 64$) or equivalently $2^5 = 32$ energy levels. For the simulation, we can fix $\omega_x = \omega_z = 1$, and the simulation time is chosen as $T = 10/\omega_z = 10$. Also, we choose 100 neurons and three hidden layers for the neural network architecture. In figure~\ref{Figure1}, we plot the solution of this control problem for five qubits and two independent control functions $g_0(t)$ and $g_p(t)$. We observe that the efficiency of the PINN approach in terms of the final fidelity amounts to $99.8 \%$, demonstrating that our approach works even in larger quantum systems without interaction. At the simulation time $T$, our PINN yields an expectation value $\langle H_p\rangle$ that amounts to $0.004$, which is close to the theoretical value $\min_{|\Psi\rangle} \langle H_p \rangle = 0$.

\begin{figure*}[ht]
	\centering
	\includegraphics[scale=0.5]{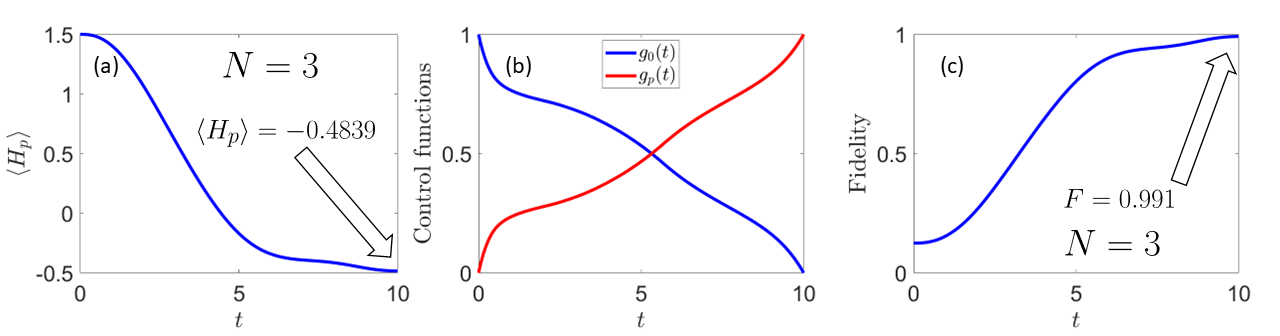}\\
	\caption{(a) Expectation value $\langle H_p \rangle$ as a function of time. (b) Control functions obtained from our PINN approach using $g_0(t) = 1 -g_p(t)$. (c) Fidelity of the protocol.}
	\label{Figure2}
\end{figure*}

\subsection{Interacting qubits}

Let us consider the following Hamiltonians

\begin{eqnarray}
H_0 &=&-\frac{\omega_x}{2}\sum_{j=1}^{N}\sigma_{x,j}, \\
H_p &=& \frac{\omega_f}{2}\sum_{j=0}^{N}(1-\sigma_{z,j}) - J \sum_{i=1}^{N-1}\sigma_{z,j}\sigma_{z,j+1}
\end{eqnarray}

For the simulation, we consider the same quantum control problem explained in the previous section but using $N=3$ qubits ($2^3 = 8$ energy levels). In this case, we use $\omega_x = \omega_f = 1$ and $J=\omega_f/4$ with a simulation time $T = 10/\omega_f = 10$. Also, we use the constraint $g_0(t) = 1 -g_p(t)$ to solve this problem.

In figure~\ref{Figure2}, we plot the solution of this control problem for three interacting qubits and two control functions $g_0(t)$ and $g_p(t)$ with the constraint $g_0(t) = 1 -g_p(t)$. We observe that the efficiency of the PINN approach in terms of the final fidelity amounts to $99.1 \%$. Moreover, our PINN scheme reaches an expectation $\langle H_p \rangle$ value of $-0.4839$, close to the theoretical value $\min_{|\Psi\rangle} \langle H_p \rangle = -0.5$. These results demonstrate that our approach performs well in larger quantum systems with interactions.

\section{Summary of PINNs parameters}

To summarize our results concerning the application of PINNs for quantum control, in Table~\ref{Table1}, we show the relevant information about how we implemented our method for different systems. In particular, we show the value for the hidden layers, epochs, neurons, control ($\eta$) and regularization ($\chi$) weights, and learning rate ($\lambda$).

\begin{table*}
\centering
\caption{\label{Table1}The Table displays relevant information about the PINNs. Our PINNs are feed-forward neural networks with hard constraints on the initial conditions. $\eta$ and $\chi$ stand for the control and regularization weights, respectively. $\lambda$ accounts for the learning rate.}

\centering
\begin{tabular}{|c|c|c|c|c|c|c|c|} 
\hline
System  & Hidden Layers  & Epochs & Neurons & $\eta$ & $\chi$ & $\lambda$   \\
\hline
\includegraphics[width=0.18\linewidth]{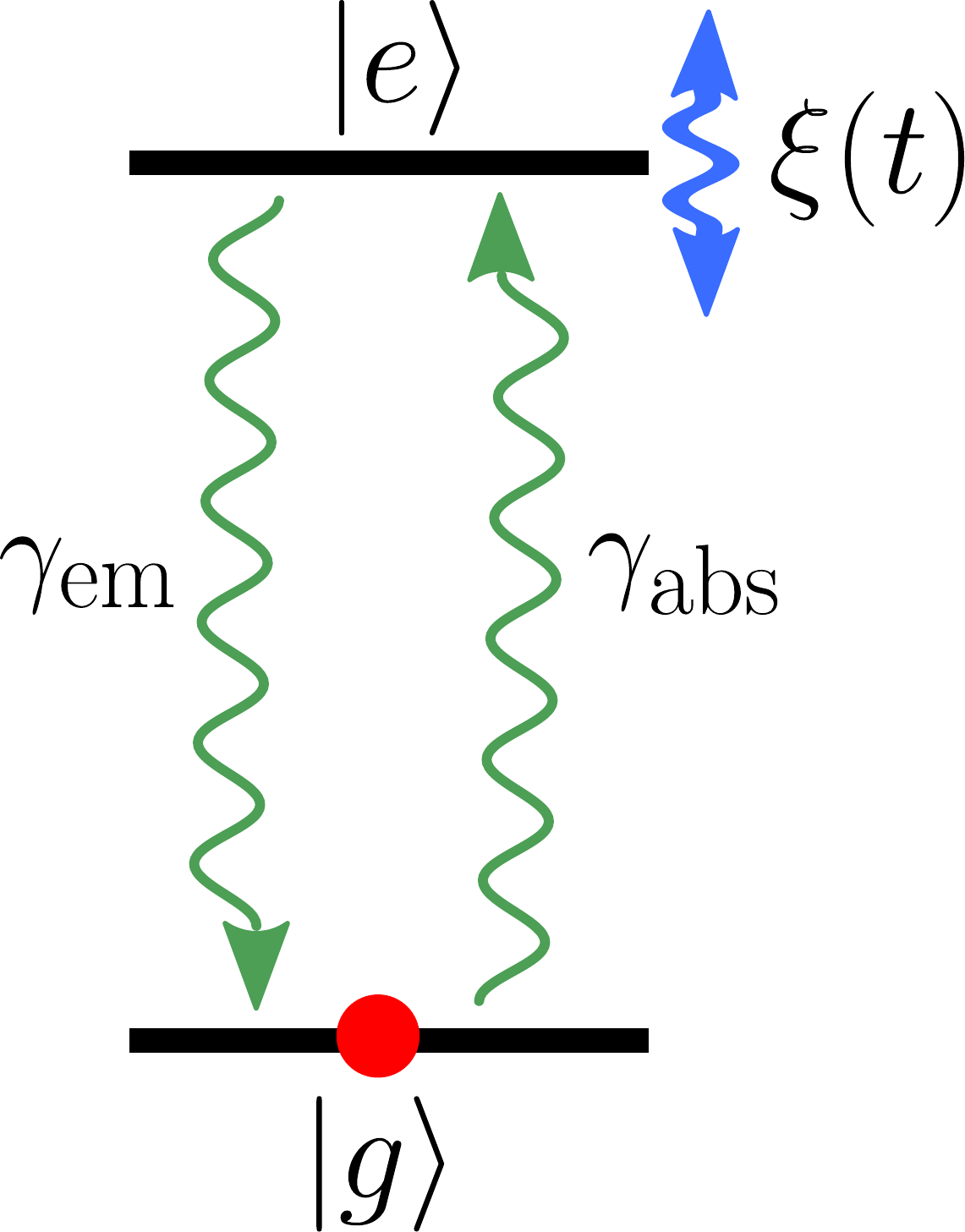} &  $4$   & $4\times10^4$ & $200$ & $0.1$ & $10^{-3}$ & $10^{-4}$  \\
\hline
\includegraphics[width=0.25\linewidth]{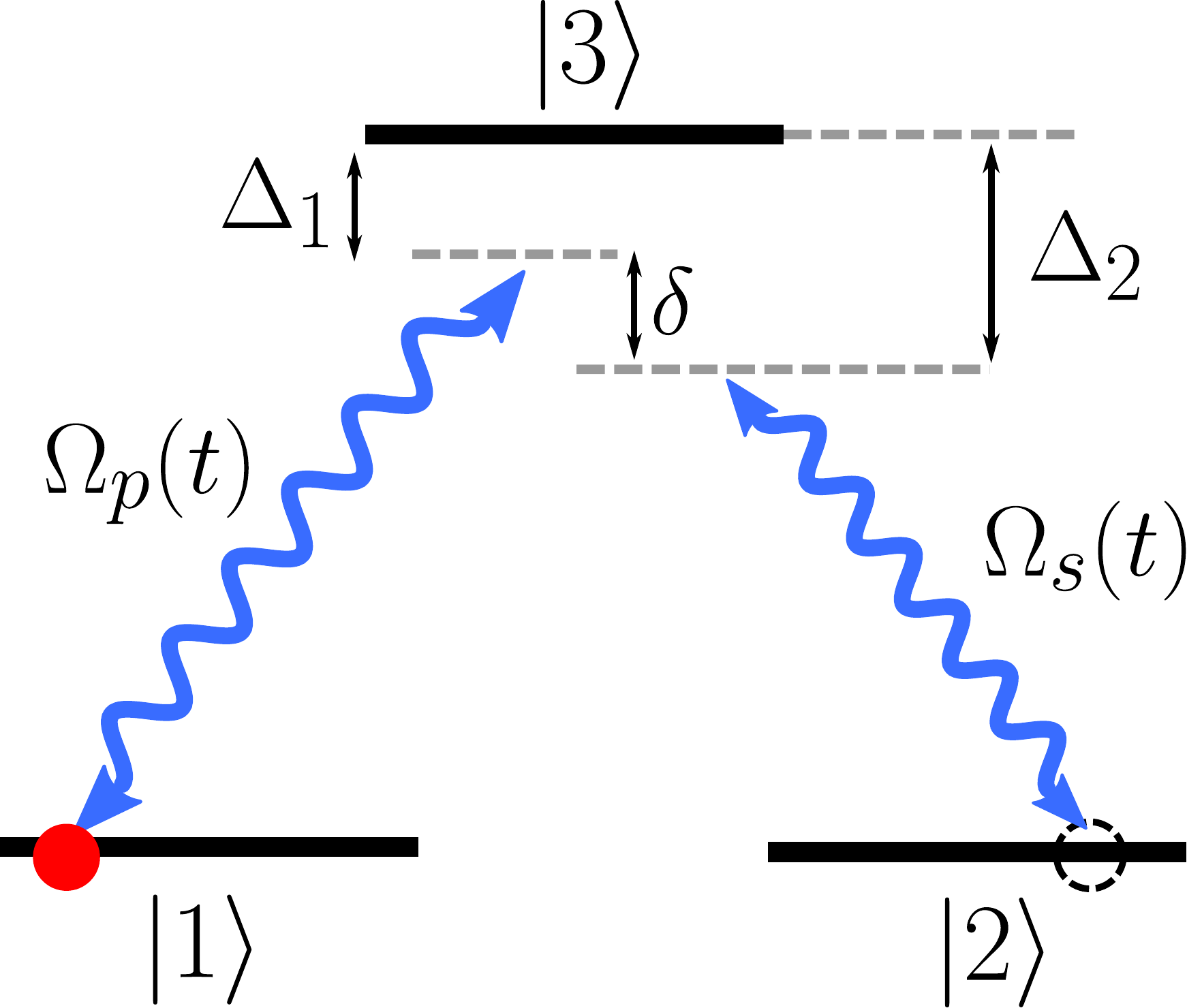} & $5$   & $2\times 10^{4}$ & $150$ & $0.2$ & $2.8\times 10^{-3}$ & $8\times 10^{-3}$ \\
\hline
\includegraphics[width=0.36\linewidth]{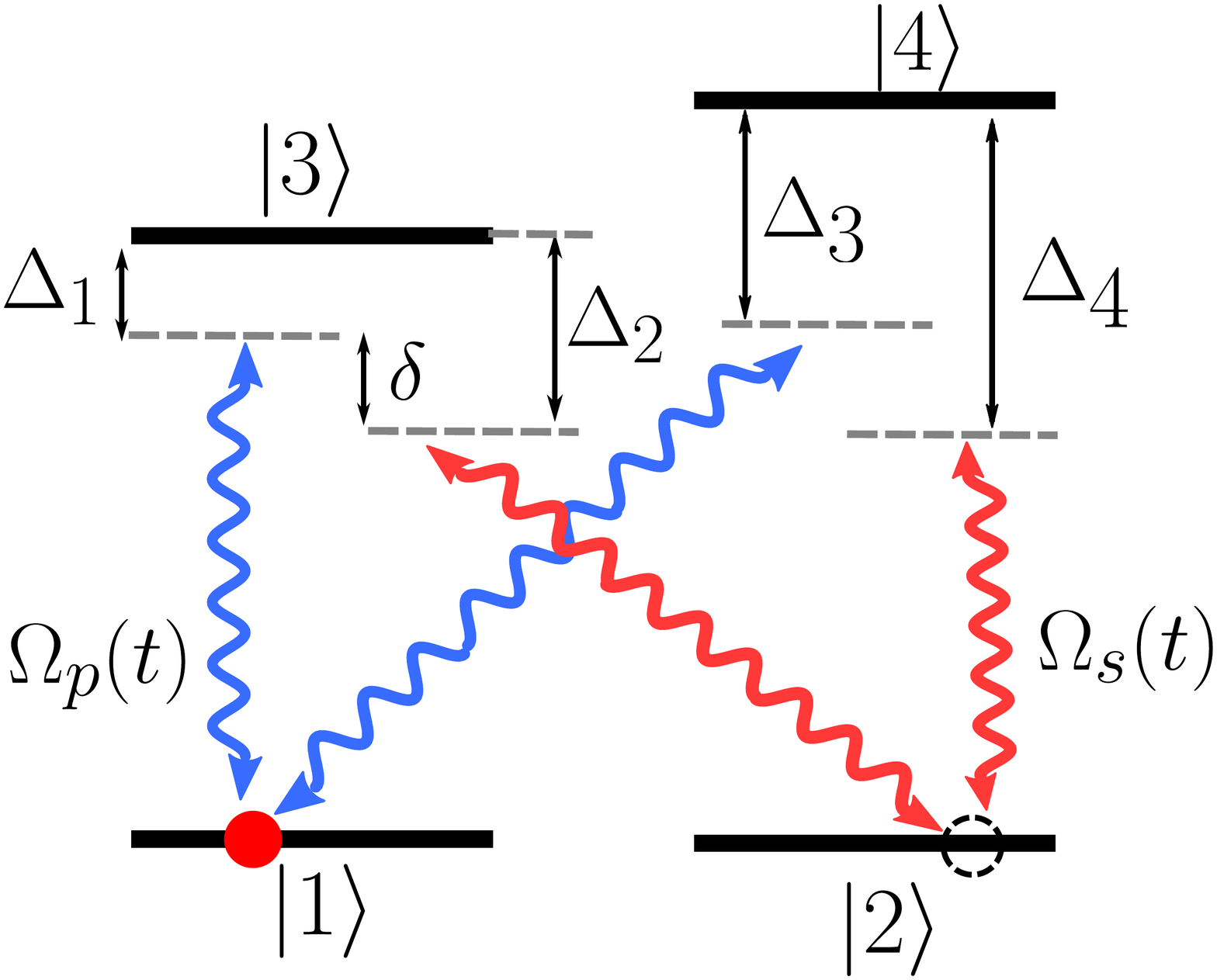} & $5$   & $2\times10^{4}$ & $150$ & $0.2$ & $2.8\times 10^{-3}$ & $8\times 10^{-3}$ \\
\hline
\includegraphics[width=0.2\linewidth]{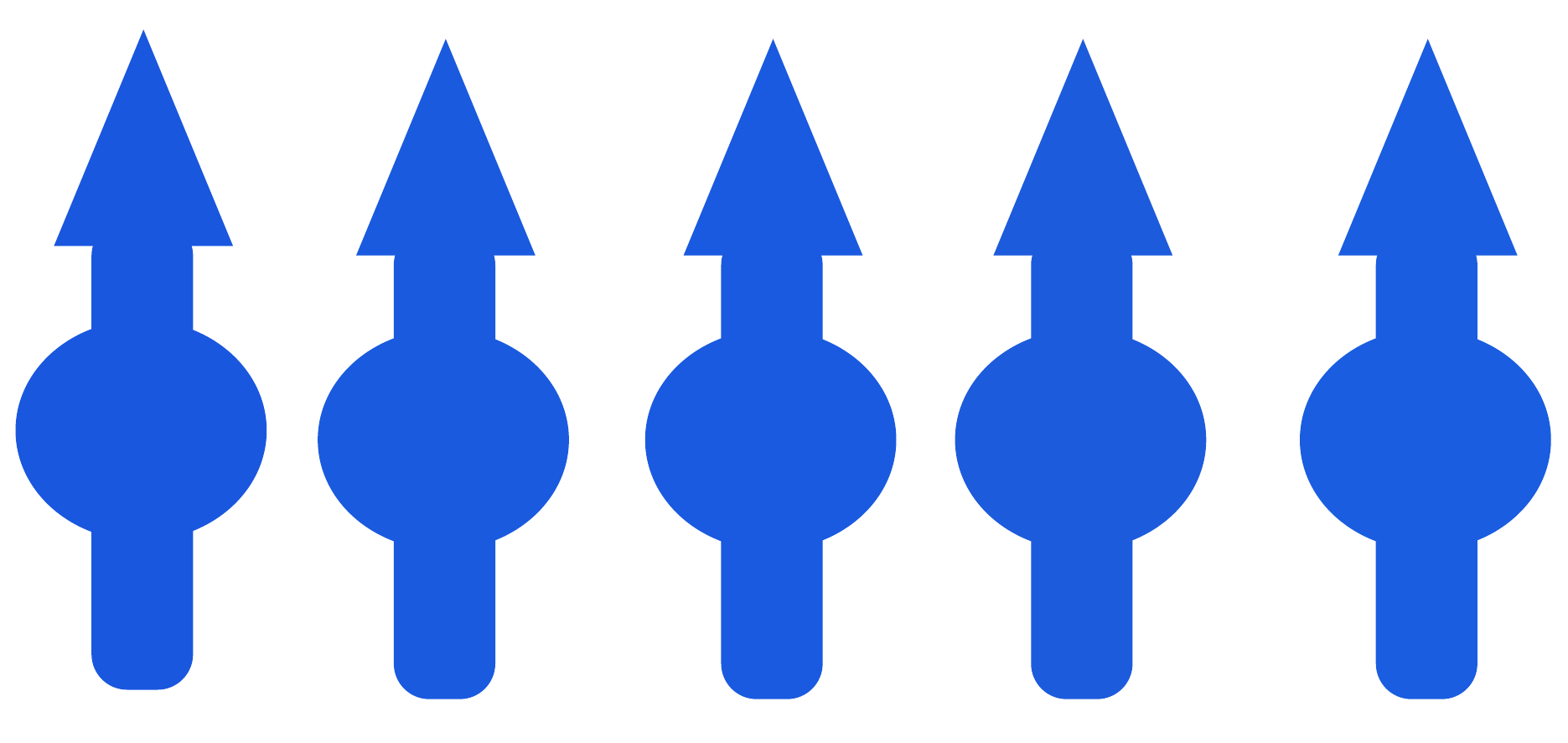} & $1$   & $10^{5}$ & $3$ & $0.05$ & $10^{-5}$ & $10^{-3}$ \\
\hline
\includegraphics[width=0.2\linewidth]{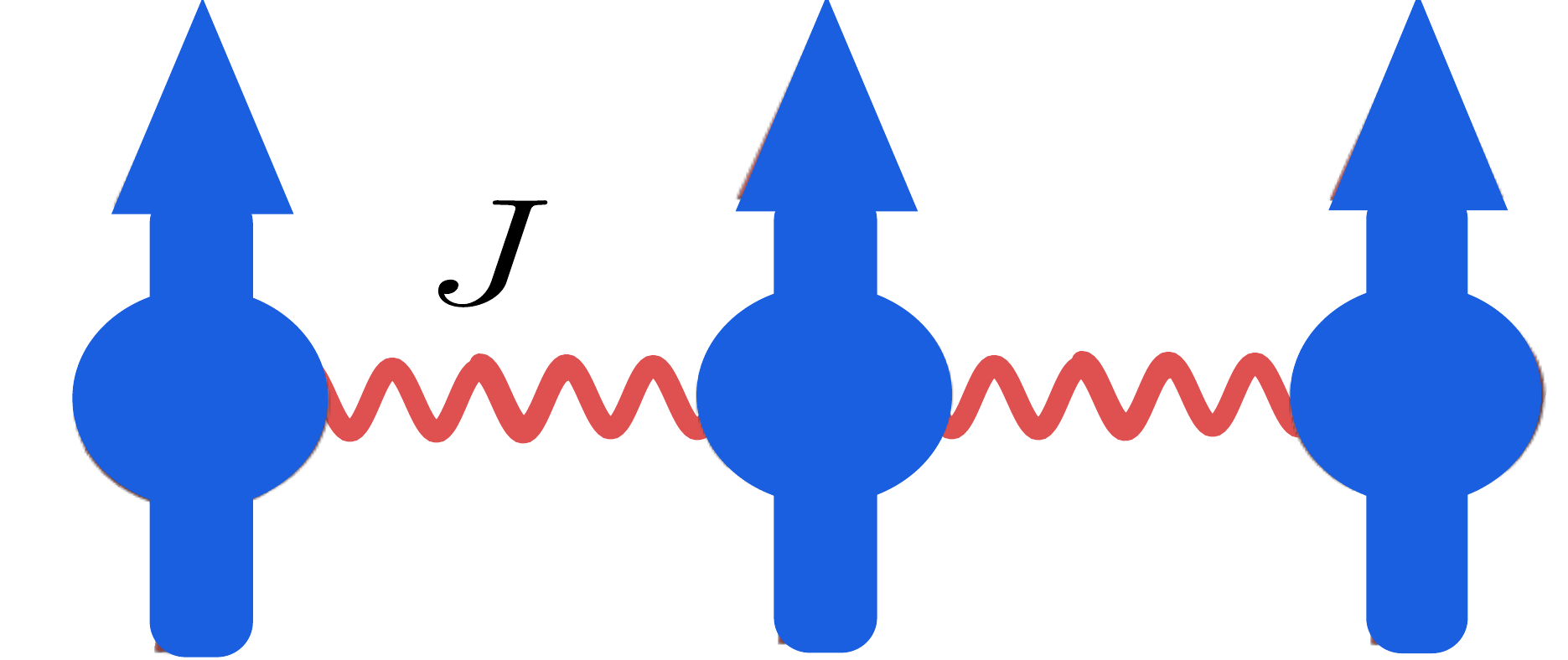} & $4$   & $5 \times 10^5$ & $100$ & $0.01$ & $10^{-4}$ & $10^{-3}$ \\
\hline
\end{tabular}
\end{table*}

\end{document}